\documentclass{aa}  

\usepackage{graphicx}
\usepackage{txfonts}
\usepackage{xcolor}
\usepackage{ulem}
\usepackage{multicol}
\usepackage{mathtools}
\usepackage{amsmath}
\usepackage[toc]{appendix}
\newcommand{\afe}{[$\alpha$/Fe]}
\newcommand{\feh}{[Fe/H]}
\newcommand{\ms}{M$_{\odot}$}

\newcommand{\teff}{$T_\mathrm{eff}$}
\newcommand{\teffzams}{$T_\mathrm{eff}^\mathrm{ZAMS}$}

\begin{document} 

   \title{A coherent view of Li depletion and angular momentum transport to explain the Li plateau -- from Population~II to Population~I stars} 

   \authorrunning{Sviatoslav~Borisov et al.}
   \titlerunning{Lithium depletion -- from Pop~II to Pop~I stars}

   \author{Sviatoslav~Borisov
          \inst{1},
          Corinne~Charbonnel,
          \inst{1,2}
          Nikos~Prantzos,
          \inst{3}
          Thibaut~Dumont,
          \inst{4}
          Ana~Palacios
          \inst{5}
          }

    \institute{Department of Astronomy, University of Geneva, Chemin Pegasi~51, 1290 Versoix, Switzerland\\
              \email{sviatoslav.borisov@unige.ch}
         \and
            IRAP, CNRS UMR 5277 \& Universit\'{e} de Toulouse, 14 avenue Edouard Belin, 31400 Toulouse, France
         \and
            Institut d’Astrophysique de Paris, UMR7095 CNRS, Sorbonne Universit\'{e}, 98bis Bd. Arago, 75104 Paris, France
         \and
            University of Strasbourg, CNRS, IPHC UMR 7178, F-67000 Strasbourg, France
         \and
            LUPM, Universit\'{e} de Montpellier, CNRS, Place Eugène Bataillon, 34095 Montpellier, France
             }

   \date{Received September 15, 1996; accepted March 16, 1997}

 
\abstract{Unraveling the cosmological Li problem -- the discrepancy between Big Bang nucleosynthesis predictions and observed values in the Spite plateau -- requires a comprehensive exploration of stellar evolution. In this study, we utilized the code STAREVOL to compute the stellar evolution models with atomic diffusion, rotation-induced processes, parametric turbulence, and additional viscosity. We calibrated the models to fit the abundance of Li in Population~II stars selected from the GALAH~DR3 spectroscopic survey and literature compilation based on their chemical composition. The calibration reveals the significance of parametric turbulence in counteracting atomic diffusion effects. These models predict the constancy of the Spite plateau as a function of \teff ~and \feh \ which agrees with the observational trend found after a detailed selection of dwarf non-peculiar stars. Other dwarfs that lie below the Spite plateau are either CEMP or have other types of chemical peculiarities, reinforcing the notion of their environmental origin. The Li abundance near the Spite plateau of the most Fe-deficient star, J0023+0307, which is not CEMP, provides additional evidence for the stellar depletion solution of the Li cosmological problem. Also, our models predict a transition from Li constancy at low metallicities to dispersion at high metallicities which is seen in observations. In addition, we extend our analysis to include a comparison with observational data from the globular cluster NGC~6752, showcasing excellent agreement between model predictions and Li and Mg trends in post-turnoff stars. This opens avenues for refining the estimates of initial Li abundance in metal-rich globular clusters which would help to constrain Li evolution in the Milky Way.} 


   \maketitle
%


\section{Introduction}
\label{sec:intro}

The heaviest and most abundant lithium isotope ($^7$Li, hereafter Li) is known to be the only one in the periodic table that has three different astrophysical sources:  thermonuclear reactions in the expanding universe during the first minutes after the Big Bang (so-called Big Bang nucleosynthesis, or BBN), spallation nuclear reactions between Galactic cosmic rays (GCR) and interstellar matter, and thermonuclear reactions in stars. The first of the three sources of $^7$Li is theoretically predicted by BBN models \citep{Wagoner1967} and observationally confirmed by the discovery of the so-called ``Spite plateau'' \citep{Spite1982}. The second one is theoretically predicted by \cite{Reeves1970} and observationally inferred by the fact that the minor isotope $^6$Li can only be produced in GCR \citep{Meneguzzi1971}. The contribution of those two sources together has been quantitatively evaluated to less than half of the solar abundance of $^7$Li \citep{Prantzos2012}; this makes the stellar source the most important Li producer in the Universe. However, although several stellar sites of $^7$Li production have been identified over the years (novae, Red Giants, Asymptotic Giant Branch stars, core-collapse supernovae and even merging white dwarfs) and despite decades of theoretical and observational efforts, the major contributor (assuming there is one) has not been identified yet. The reasons are both theoretical and observational.

Indeed, the yields of the various candidate sources are still highly uncertain. This concerns stars on the red giant branch \citep[RGB,][]{Sackmann1999}, the asymptotic giant branch  \citep[AGB,][]{Sackmann1992,2009PASA...26..145I,vanRaai2012,Karakas2016}, classical novae \citep[][]{1975A&A....42...55A,1978ApJ...222..600S,2020ApJ...895...70S,Jose1998,Jose2006,Denissenkov2014}, core-collapse supernovae \citep{Woosley1990,Woosley1995,2010IAUS..268..463N,2019ApJ...872..164K}, or merging white dwarfs \citep{2012A&A...542A.117L}. This limits the predictive power of Galactic chemical evolution models \citep{Travaglio2001,Prantzos2012,2021MNRAS.505..200M,2021A&A...656A..64R,2021A&A...653A..72R,2022ApJ...933L..30K,2023arXiv230112299G}. Interest in novae has been revived in recent years, through observations and quantitative evaluation of their Li yields \citep[see][and references therein]{Molaro2023}. However, the uncertainties affecting the evolution of the nova rate, as well as the overproduction of minor CNO isotopes by nova models {\it if} novae are the major Li producers, still prevent a definitive conclusion \citep{Jose1998,Prantzos2017}.

On the other hand, the theoretical contribution of BBN (so-called primordial Li) is constrained to a very high degree of accuracy in the standard cosmological model from the analysis of the Cosmic Microwave Background (CMB, \citealt{2020A&A...641A...6P,2021A&A...652C...4P}) and nuclear and particle physics \citep[e.g.][]{2007ARNPS..57..463S,2016RvMP...88a5004C,Pitrou2018,2020ApJ...901..127I,2021ApJ...915L..13H,2023JPhCS2619a2012P}. However, the primordial Li/H value predicted by BBN+CMB exceeds by a factor of $\sim$three the so-called Li plateau on which the highest photometric Li abundances observed in old, metal-poor Galactic turnoff stars, which are the only observable relics of the early Universe for this isotope, are lying \citep{1982Natur.297..483S,1988A&A...192..192R,1994ApJ...421..318T,2001ApJ...549...55R,Charbonnel2005,Sbordone2010,2010A&A...511A..47H,2020MNRAS.497L..30G}.  The Li plateau extends over a wide range in [Fe/H] (up to -1.5~dex, with a possible meltdown in the very Fe-poor regime below $\sim-3.0$~dex; e.g., \citet{2017AJ....154...52M} and references therein) and a specific range in effective temperature (above $\sim 5800-6000$~K depending on the observational studies; e.g., \citet{Norris2023} (hereafter N23) and references therein). A similar Li plateau has been discovered in dwarf spheroidal and ultra-faint galaxies, as well as in accreted satellites such as Gaia-Enceladus and three members of the S2-Stream  \citep{2014MNRAS.444.1812M,2019A&A...626A..15H,2020MNRAS.496.2902M,2021MNRAS.500..889A,2021MNRAS.505..200M}. It has also been found in $\omega$~Centauri \citep{Monaco2010}. This pattern thus does not depend on environmental effects, even though the lack of stars with Li plateau values in the extremely Fe-poor regime ([Fe/H] $\leq-4$~dex) might require explanations. It has been suggested that the astration of a large fraction of the interstellar medium (ISM) of galactic haloes by a first generation of stars would have effectively destroyed Li before the formation of the halo dwarfs we are observing today \citep{Piau2006}, but this would lead to an important overproduction of metals \citep{Prantzos2010}. Alternatively, a strong primordial magnetic field could lead to the chemical separation of Li+ ions during the hierarchical structure formation and reduce its abundance within the collapsed structures \citep{2019ApJ...876L..30K}. Physics beyond the Standard $\Lambda$CDM model has also been invoked to lower the theoretical primordial Li abundance and reconcile it with the level of the Spite plateau \citep{2004PhRvD..70f3524J, 2009PhR...472....1I,2011ARNPS..61...47F,2009PhR...472....1I,2011ARNPS..61...47F,Coc2017,DealMartins2021}. In this case, however, the challenge is to keep the predictions for the primordial abundances of deuterium and helium-4 unaffected, since the present agreement with those deduced from the absorption spectra of high-redshift quasars and observations of HII regions in metal-poor galaxies, respectively, is excellent \citep{2014MNRAS.445..778I,2021MNRAS.505.3624V,2016ApJ...830..148C,2018MNRAS.477.5536Z,2020ApJ...896...77H,2021JCAP...03..027A}, making the concordance between CMB and BBN one of the greatest triumphs of physics. 

While Beyond Standard Model physics remains open  \citep[][and references therein]{2023arXiv230112299G}, many studies have explored the so-called stellar depletion solution to reconcile the Li value predicted by standard BBN and its abundance along the plateau, considering that this fragile element burns at relatively low temperature ($\sim2.5\times10^6$~K) in stellar interiors. This is supported by decades of spectroscopic surveys that have depicted the Li abundances in low-mass stars belonging to different Galactic subpopulations (halo, open and globular clusters, thin and thick discs). It is now firmly established that the photospheric abundance of Li decreases as low-mass stars (including the Sun) age on the pre-main and main sequences, in proportions that depend on several factors among which the mass and the metallicity of the star, and its rotation rate \citep[e.g.][]{1993AJ....106.1059S,2005A&A...442..615S,2018A&A...615A.151B,2019AJ....158..163D,2021MNRAS.500.1158J}. It is also well accepted that microscopic (i.e., atomic diffusion) and macroscopic transport processes easily modify the chemical element distribution within stellar interiors as well as the surface composition compared to that with which low-mass stars were born \citep[e.g.][and references therein]{2015ads..book.....M}. Clear signatures of their interactions and global effects were evidenced through the observed trends  with evolutionary stage of the atmospheric abundances of various heavy elements in Population II turnoff and subgiant stars in several globular clusters \citep{2006Natur.442..657K,2007ApJ...671..402K,2012ApJ...753...48N,Gruyters2013,2016A&A...589A..61G,2021A&A...652A..75G}. In their seminal paper, \citet{2002ApJ...580.1100R} concluded that ``the Spite plateau for Li in low-metallicity field stars remains the strongest argument for the presence of a (hydrodynamic) process competing with atomic diffusion''. These macroscopic processes can also lead to the photospheric Li depletion in Population I as well as in Population II low-mass stars \citep[Pop I and PopII hereafter; e.g.][]{1984ApJ...282..206M,1988ApJ...335..971V,Proffitt1991,1994ApJ...433..510C,1995A&A...295..715V,1998ApJ...502..372V,2000astro.ph..6279P,2001A&A...376..955S,2002ApJ...568..979R,Richard2005,2006Natur.442..657K,2008ApJ...689.1279P,2013A&A...552A.131V,2020A&A...633A..23D}.  However, while atomic diffusion can be computed from first principles as it results from internal gradients of pressure, temperature, and composition, the description of macroscopic flows of different natures (turbulence, rotational currents, magnetic instabilities, convection, mass loss, accretion) and of their interactions remains extremely challenging in 1D stellar evolution models. The fact that some of these mechanisms transport both chemical elements and angular momentum in stellar interiors adds another layer of complexity, leading in some cases to the use of simple parameterizations that can then constrain the underlying physics \citep[e.g.][]{2006ApJ...645..634T,2009A&A...495..271D,2013LNP...865...23M,Dumont2021a}, with additional constraints from asteroseismology in the case of Population I stars \citep[e.g.][]{1996A&A...312.1000R,2003A&A...408.1037P,2010ApJ...713.1108G,2021A&A...647A.122M,2022NatAs...6..788E,2023A&A...669L...9B}. 

The actual challenge in reproducing the Li plateau lies in its constancy over a wide range in [Fe/H] and a specific range in effective temperature for Pop~II stars, whereas Li varies greatly among Pop~I stars. This is what we explore in this work, applying in stellar models of low-metallicity stars along the Li plateau the same assumptions for the transport of chemicals and angular momentum as \citet[][hereafter D21a,b]{Dumont2021b,Dumont2021a} who studied lithium and rotation in solar-type field stars and F- and G-type stars in Galactic open clusters. This paper is structured as follows: in \S\ref{sec:stellar_models}, we discuss the input physics and transport processes included in our models. \S\ref{sec:calibration} is dedicated to the calibration of the models to fit the Li plateau value. We discuss in \S\ref{sec:obs_data} data selection and properties of the observed field Pop~II stars and how they match the model prediction. In this section, we also discuss the reasons for the dispersion of Li abundance at high metallicities. In addition, we also compare our models with the behaviour of Li and Mg in post-turnoff stars of the globular cluster NGC~6752. We summarize our results in \S\ref{sec:summary}.

\section{Stellar evolution models}
\label{sec:stellar_models}
\subsection{Basic input physics of the grid of models}
\label{subsec:input_physics}
For stellar evolution modeling, we use the code STAREVOL \citep[][]{Siess2000,Palacios2006,Decressin2009,Lagarde2012,Amard2019} with the same basic input physics and prescriptions for various transport processes of chemicals and angular momentum (see \S\ref{subsec:transport}) as in D21b. The models are computed with the OPAL opacity tables \citep{Iglesias1996} and Wichita opacity database (Ferguson, J.~W., private communication) for the high- and low-temperature regimes ($T>8000$~K and $T<8000$~K respectively), nuclear reaction rates from the NACRE2 database \citep{Xu2013a,Xu2013b}, and the analytical equation of state of \citet{Eggleton1973} and \citet{Pols1995} described in \citet{Siess2000}. The surface boundary conditions account for  \citet{KrishnaSwamy1966} correction to the grey atmosphere approximation with the numerical surface and the connection to the atmosphere set at optical depths equal to $5\times 10^{-3}$ and 2 respectively \citep[for details see][]{Amard2019}. The convective boundaries are obtained with the Schwarzschild criterion, with the value for the mixing length parameter ($\alpha_{MLT}= 2.2236$) that was calibrated on the solar model with the same input physics including atomic diffusion and rotation by D21b. We adopt the corresponding initial mass fraction Y for He in the Sun (0.2718). For models of different metallicities (Z, mass fraction of metals heavier than He), we assume the $\Delta Y / \Delta Z$ value derived from the above-mentioned solar calibration with $Y_0=0.2484$ \citep{Coc2017} and assume the solar chemical mixture from \citet{Asplund2009} with Ne modified as suggested by \citet{Young2018}.

\begin{figure}
\includegraphics[width=1\linewidth]{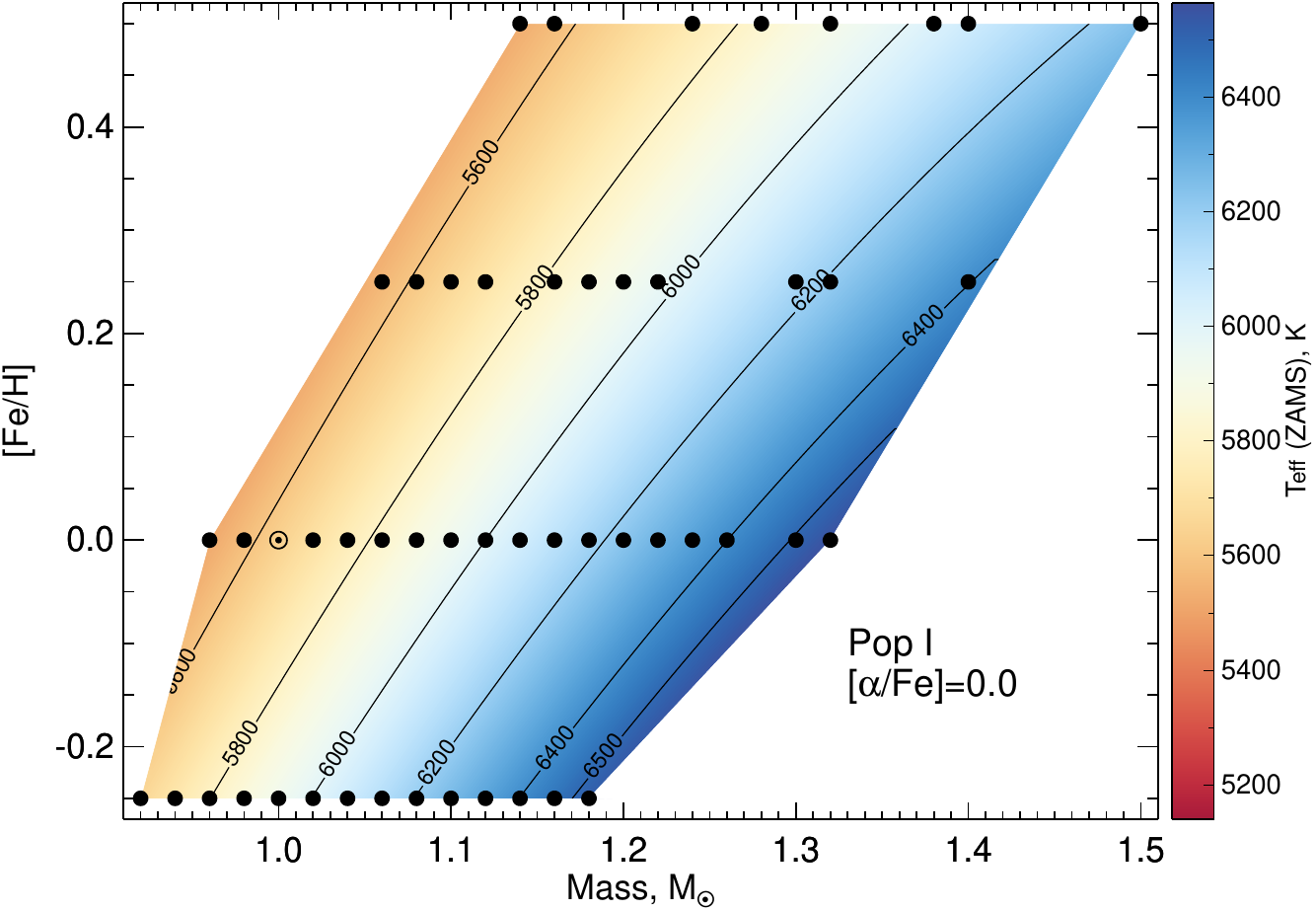}
\includegraphics[width=1\linewidth]{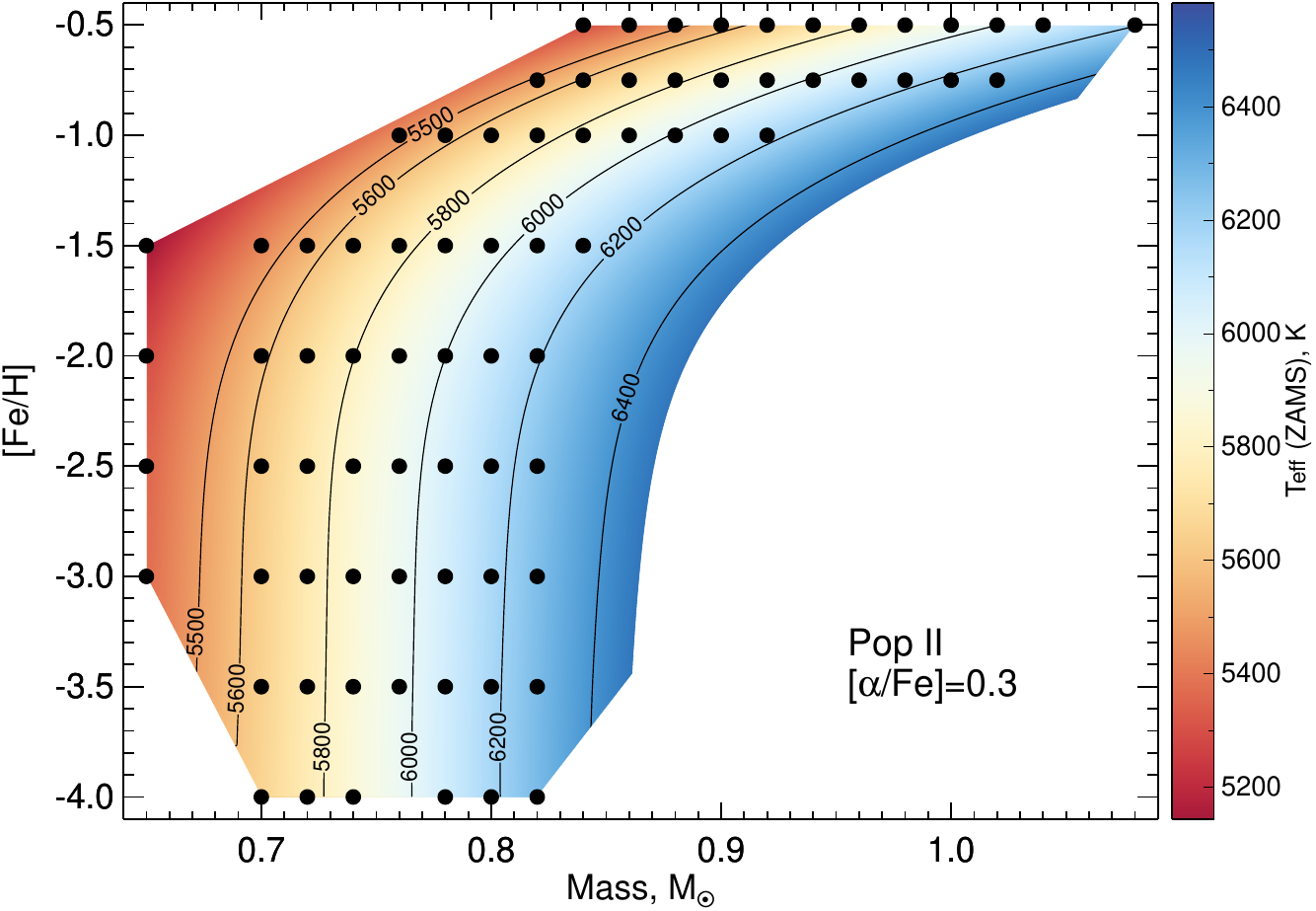}
\caption{The model grid computed for Pop~I and Pop~II stars down to [Fe/H]=-4~dex (upper and bottom panels respectively). The circles (including the solar symbol) indicate the masses and \feh ~values of the computed models, whilst the background shows the predicted values of \teffzams \ which have been interpolated with a third-degree polynomial.} 
\label{grid_teff_zams}
\end{figure}

With this input physics, we computed two grids of models that are illustrated in Fig.~\ref{grid_teff_zams}:
\begin{itemize}
    \item a grid of Pop~II models with [Fe/H] between $-4.0$ and $-0.5$~dex and [$\alpha$/Fe] enhancement by 0.3~dex, with an extension down to -5.8~dex with tailor-made models for the most Fe-poor star of our sample, J0023$+$0307 (\S~\ref{selection-field} and \ref{subsec:comp_obs_data}). For this grid, we assume an initial A(Li) equal to the primordial BBN prediction A(Li)=2.75~dex \citep{Pitrou2018,Coc2017}. While Pop~II stars have [Fe/H] lower than -1.5 or -1~dex (depending on studies), we intentionally add models up to [Fe/H]=-0.5~dex for comparison purposes with the Pop~I models. The mass range covered by this grid corresponds to stars presently (i.e., at an age of $\sim$12~Gyr) lying in the \teff ~range of the Spite plateau (most stars have \teff \ between $\sim$5800~K and 6500~K) as illustrated on the lower panel of Fig.~\ref{grid_teff_zams} (we show only models with [Fe/H] down to -4, as the iso-\teff \ lines stay vertical below that value). \\
    \item a grid of Pop~I models with [Fe/H] between  $-0.25$ and $<+0.5$~dex and [$\alpha$/Fe]=0 with initial masses chosen so that the models had the same effective temperature on the zero-age main sequence (ZAMS) as the Pop~II stars of the plateau. This is illustrated in the upper panel of Fig.~\ref{grid_teff_zams} with color-coded \teffzams. For these models, we assume a linear increase of initial Li with Fe-content up to the meteoritic Li value of 3.31~dex \citep{Anders1989} at the metallicity of the Sun.
\end{itemize}

All the models are computed from the pre-main sequence including the transport processes described below. 

\subsection{Transport processes of angular momentum and chemicals}
\label{subsec:transport}
\subsubsection{Global picture}
\label{subsec:transportglobal}

Consistently adopting the nomenclature established by D21a,b, we specifically focus on their $_{\nu}R1^{T_0}_A$ model\footnote{In the adopted nomenclature $_{\nu}R1^{T_0}_A$, $\nu$ stands for the ad hoc vertical viscosity $\nu_{add}$, $T_0$ is the logarithm of the temperature at which the parametric turbulence is anchored, and $A$ stands for the prescription for penetrative convection.} that they calibrated to fit the observed abundance of Li of stars around solar metallicity and the radial rotation profile of the Sun. It includes the transport of angular momentum driven by meridional circulation, shear-induced turbulence, and parametric vertical viscosity, as well as the transport of chemicals by atomic diffusion, rotation-induced processes, penetrative convection, and parametric turbulence, with the prescriptions we recall below. The corresponding references are summarized in Table~\ref{table_processes} together with the values of the key parameters that were adjusted by D21a,b for field solar-type stars, G- and F-type stars, with the exception of stars in the Li-dip\footnote{It was shown by D21a that a different prescription for the vertical shear was required to reproduce the Li-dip that is observed in a group of F-type stars centered around $\approx$6600~K, as observed e.g. in the Hyades and Praesepe open clusters \citep{Cummings2017}.}, in open clusters with [Fe/H] between -0.4 and +0.15~dex. 

The evolution of the abundances in the models is described by the following equation 
\begin{equation}
    \rho \frac{\partial X_i}{\partial t} = \frac{1}{r^2}\frac{\partial}{\partial r} \left(r^2 \rho \left(D \frac{\partial X_i}{\partial r}-X_i \mathrm{v}_i\right)\right)+m_i\left[\sum_jr_{ji}-\sum_kr_{ik}\right],
    \label{gen_eq}
\end{equation}
where $\rho$ and $r$ are the density and the radius, $X_i$ is the mass fraction of the nuclei $i$ of mass $m_i$ and of atomic diffusion velocity v$_i$, $r_{ji}$ and $r_{ik}$ are the rates of the nuclear reactions producing and destructing nuclei $i$ from nuclei $j$ and $k$ respectively. Finally, $D$ is the sum of all the considered diffusion coefficients that describe rotation-induced mixing (meridional circulation and shear-induced turbulence), penetrative convection, and parametric turbulence as described below.

\begin{table*}[h]
\centering
\caption{Transport processes included in our models.}
\begin{tabular}{lccc} 
\hline
\hline
Process [Quantity]    &   Reference               &   Parameters adjusted &  \\
 & & for Pop~I (D21ab) & for Pop~II \\
\hline
Atomic diffusion [C]   & \citet{Thoul1994,Paquette1986} &   --  & \\
Parametric turbulence $\mathrm{D_{T}}$ [C] & \citet{Richer2000, Richard2005} &   log$T_0$=6.42 & log$T_0$=6.28\\
Penetrative convection $\mathrm{D_{A}}$ [C]  &   \cite{Augustson2019}    &   $d_\mathrm{ov}=0.0325$ 
& \\ 
Horizontal shear $\mathrm{D_{h}} - \mathrm{\nu_{h}}$ [C-AM]     &   \citet{Mathis2018}      &   --   &   \\
Vertical shear $\mathrm{D_{v}} - \mathrm{\nu_{v}}$ [C-AM]  &   \citet{Zahn1992}        &   --   &   \\
Parametric viscosity $\mathrm{\nu_{add}}$ [AM] & \citet{Eggenberger2012}; D21a & $\nu_{add}$=3.5$\times$10$^4$~cm$^2$ s$^{-1}$ & \\
Magnetic torque [AM]   &   \citet{Matt2015, Amard2019}    & 
K=7.5$\times$10$^{30}$ & \\
\hline
\end{tabular}
\tablefoot{C for chemicals and AM for angular momentum.}
\label{table_processes}
\end{table*}

\subsubsection{Transport of angular momentum and rotation-induced mixing}
\label{subsubsec:transport_am}
The treatment of stellar rotation in our models follows the formalism developed by \citet{Zahn1992}, \citet{Maeder1998}, and \citet{Mathis2004} under the so-called shellular rotation hypothesis. The transport of angular momentum obeys the following advection-diffusion equation:
\begin{equation}
    \rho \frac{d}{dt} (r^2 \Omega)=\frac{1}{5r^2}\frac{\partial}{\partial r}\left(\rho r^4 \Omega U_2\right) +\frac{1}{r^2}\frac{\partial}{\partial r}\left((\nu_\mathrm{v}+\nu_{add}) r^4 \frac{\partial \Omega}{\partial r}\right),
    \label{transport_am}
\end{equation}
where $\rho$, $r$, $\Omega$, $U_2$, and $\nu_\mathrm{v}$ are the density, radius, angular velocity, meridional circulation velocity on an isobar, and vertical shellular component of the turbulent viscosity, respectively. For the chemicals, the transport by meridional circulation is described with the diffusion coefficient $\mathrm{D_{eff}}=\left|rU_2(r)\right|^2/(30D_h)$ entering Eq.~\ref{gen_eq} \citep{Chaboyer1992}, with D$_h$ the horizontal diffusion coefficient. As in the R1 models of D21a,b, we use the prescriptions of \citet{Mathis2018} and \citet{Zahn1992} for horizontal and vertical diffusivities D$_h$ and D$_v$ respectively and assume a proportionality factor of 1 with the horizontal and vertical viscosities. For the parametric vertical viscosity $\nu_{add}$ that was originally introduced by \citet{Eggenberger2012} in order to flatten the internal rotation profile as evidenced by helio- and asteroseismology of low-mass stars, we use the same value of $3.5\times10^4$ cm$^2$s$^{-1}$ as calibrated on the solar rotation profile by D21a,b and discuss the impact of this parameter in \S\ref{subsec:viscosity}.  

The extraction of angular momentum at the stellar surface due to magnetized winds is accounted for following \citet{Matt2015} formalism as described in \citet{Amard2019} with the parameters $m$=0.22 and $p$=2.1, which refer respectively to an exponent related to the magnetic field geometry and the exponent relating rotation and activity. For the parameter $K$ that is linked to magnetized wind braking, we use the value of $7.5\times10^{30}$~erg that was calibrated by D21a to reproduce the solar surface rotation at the age of the Sun. We assume the same initial rotation rates on the PMS as observed in open clusters \citep[][see also \citealt{Amard2016} and D21a,b]{2015A&A...577A..98G}. Median, slow, and fast rotation refer to initial rotation periods of 4.5, 9.0, and 1.6~days respectively. We set the timescale of the disc coupling at 5~Myrs in the first two cases, and 2.5~Myrs in the last one. We check the importance of these assumptions in 
\S\ref{subsec:turbulence_calib}.

\subsubsection{Penetrative convection}
\label{subsubsec:pen_conv}
Our models include penetrative convection treated as an overshoot with the formalism of \citet{Augustson2019} that affects only the chemicals. The corresponding diffusion coefficient that enters Eq.~\ref{gen_eq} is 
\begin{equation}
    D_A(r)=D_0\left[1-\exp{\left(-\exp{\left(\frac{r-r_{bcz}}{d_\mathrm{ov}\cdot(\mathrm{v/v_0})^{3/2}}+\frac{\mu}{\lambda}\right)}\right)}\right],
\end{equation}
where $D_0=\alpha_{MLT}\mathrm{v}_{conv}H_p/3$, v$_{conv}$ is the mean velocity of convective elements, $H_p=\mathrm{d}r/\mathrm{d}\ln P$ is the pressure scale height, $d_\mathrm{ov}$ is the parameter that controls the overshoot depth, and $(\mathrm{v}/\mathrm{v}_0)$ is the
ratio of the velocity of the convective elements when taking rotation into account to the non-rotating inviscid value. We used the coefficients $\mu=5\times10^{-3}$ and $\lambda=6\times10^{-3}$  prescribed by \citet{Baraffe2017}, and $d_\mathrm{ov}$=0.0325 as calibrated by D21a,b to reproduce the surface Li abundance of young open clusters. 

\subsubsection{Atomic diffusion and parametric turbulence}
\label{subsubsec:atom_diff_and_turb}

Atomic diffusion is treated with the formalism of \citet{Thoul1994}
to solve the Burgers equations and compute the individual atomic diffusion velocities of
all the isotopes that are taken into account in STAREVOL. The collision integrals are computed following \citet{Paquette1986}. We account for the partial ionization of chemical elements for temperatures lower than $5\times10^6$~K (Schlattl 2002). Our models do not include radiative accelerations, which are always smaller than gravity for Li in Population II stars along the plateau \citep[e.g.][]{2013A&A...552A.131V}.

Turbulence in the radiative zone is parametrized following the expression initially proposed by \citet{Richer2000} to compete with atomic diffusion in AmFm stars. The corresponding diffusion coefficient that enters Eq.~\ref{gen_eq} follows a simple algebraic dependence on density: 
\begin{equation}
    D_\mathrm{T}(r)=400D_\mathrm{He}(T_0)\left[\frac{\rho(r)}{\rho(T_0)}\right]^{-3},
    \label{eq_dhe}
\end{equation}
where $D_\mathrm{He}(T_0)$ is the He atomic diffusion coefficient at the layer within the star where the temperature is $T_0$ (see below), and that is computed with the analytical approximation $D_\mathrm{He}=3.3\cdot10^{-15}T^{2.5}/[4~\rho \ln(1+1.125\cdot10^{-15}T^3/\rho)]$ from \citet{Richer2000}\footnote{This approximation for $D_\mathrm{He}$ is done only when computing $D_\mathrm{T}(r)$, but not for the stellar structure computation.}. Since Li burning is very sensitive to temperature, the free parameter $T_0$ can be adjusted to lead to more or less photospheric Li depletion. The $\rho^{-3}$ dependence resulted from fitting the abundance anomalies in AmFm stars by \citet{Richer2000}. We discuss the calibration of $T_0$ and compare it to the value derived by D21a,b for Pop~I stars in \S\ref{subsec:turbulence_calib}.

\section{Model calibration}
\label{sec:calibration}

In this section, we discuss the validity of the parameters that were calibrated for Pop~I stars (see Table~\ref{table_processes}) in the models of Pop~II plateau stars for which we do not have information on their initial and internal rotation rates. We assume A(Li)=2.3~dex\footnote{A(Li)=$\log_{10}$(N$_\mathrm{Li}$/N$_\mathrm{H}$)+12 where N$_\mathrm{Li}$ and N$_\mathrm{H}$ are the number density of Lithium and Hydrogen respectively.} for the plateau value \citep{Norris2023} and 12~Gyr for the mean age of the plateau stars. We first focus on a ``reference'' star, which we chose assuming that the typical mass of the Spite stars is M=0.74\ms \ with the chemical composition of \feh=-2.0 and \afe=0.3. This star has an effective temperature of 6234~K at 12~Gyr (according to our model). We then test the robustness of the calibration across the [Fe/H] and \teff ~range of the Spite plateau.

\subsection{Additional parametric vertical viscosity}
\label{subsec:viscosity}
As extensively discussed in the literature, with the currently available prescriptions, shear turbulence and meridional circulation alone do not account for enough angular momentum transport to reproduce the internal rotation profile of the Sun and of some well-studied Pop~I main sequence stars (see references in Sect.1). As of today, there is no asteroseismic data to depict the internal rotation of Pop~II main sequence stars and to calibrate the value of the parametric additional viscosity $\nu_{add}$ for them. To the best of our knowledge, no theoretical estimate exists of the relative efficiency in Pop~II stars of the magnetic Tayler instability, which is one of the main candidates for efficient angular momentum transport in the Sun and solar-type stars \citep[][and references therein]{2022A&A...664L..16E}. On the other hand, \citet{Talon2004} showed that for Pop~II stars lying along the Li plateau, the other main candidate, namely internal gravity waves, should efficiently lead to a quasi-solid rotation state, similar to that of the Sun \citep{2002ApJ...574L.175T,2005Sci...309.2189C}.

\begin{figure}
\centering
\includegraphics[width=1\linewidth]{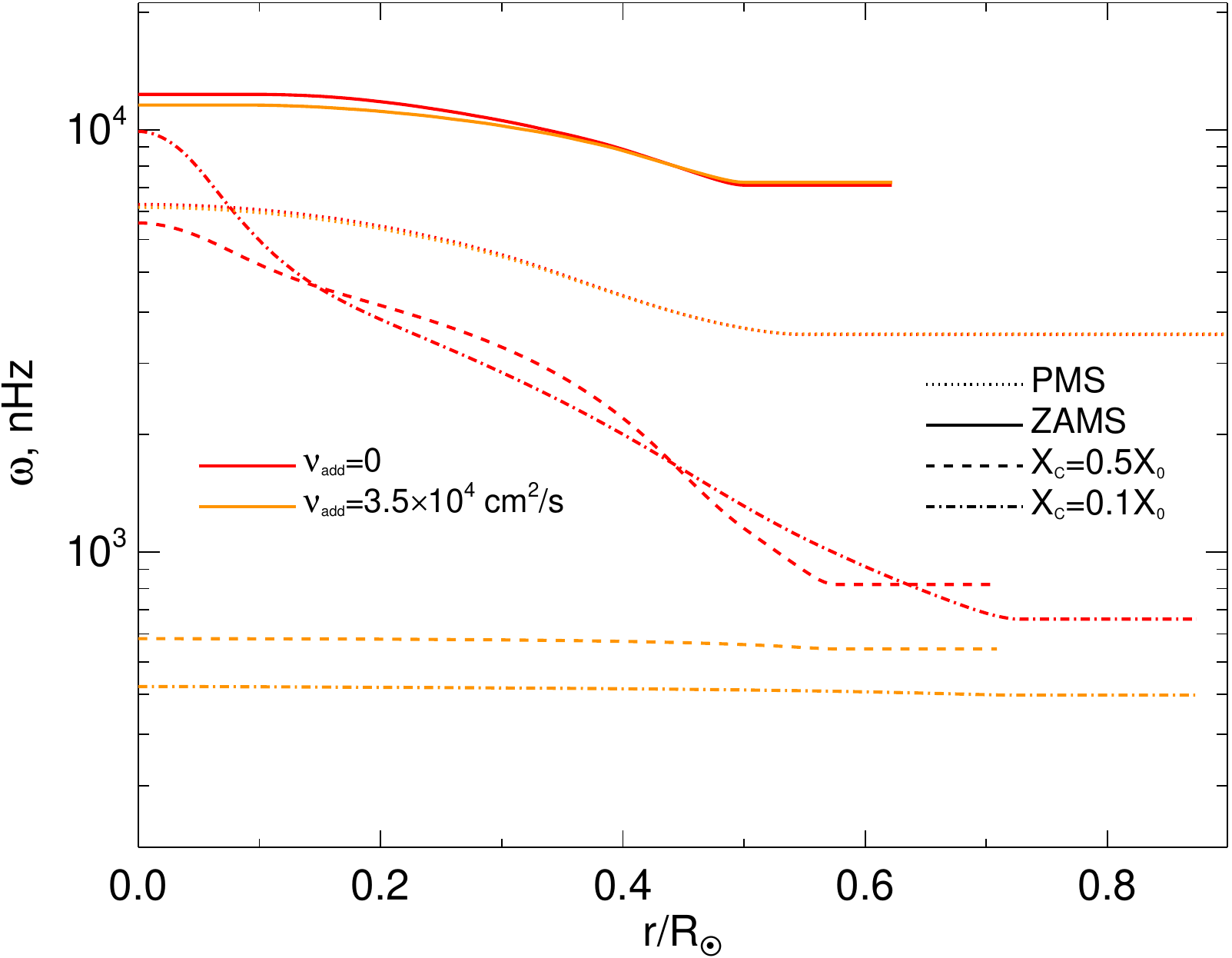}
\caption{Angular velocity profiles within the models with M=0.74\ms, \feh=-2.0, \afe=0.3 
computed with and without additional viscosity ($\nu_{add}=3.5\times10^4$ cm$^2$s$^{-1}$ and 0, red and orange lines respectively) at different evolution stages. Both models include rotation at the median rate.}
\label{angular_velocity_profiles}
\end{figure}

We assume that the theoretical behaviour of internal gravity waves legitimates the use of the same value of $\nu_{add}$ as in the R1 models of D21a,b (i.e., $3.5 \times 10^{4}$cm$^2$ s$^{-1}$, constant in space and time).  We show in Fig.~\ref{angular_velocity_profiles} the internal profiles of angular velocity at different evolution stages for two models of the reference star computed  with $\nu_{add}$=0 and $3.5 \times 10^{4}$cm$^2$ s$^{-1}$. As in the case of Pop~I low-mass stars, the additional vertical viscosity strongly flattens the internal rotation profile along the main sequence.

\begin{figure}
\centering
\includegraphics[width=1\linewidth]{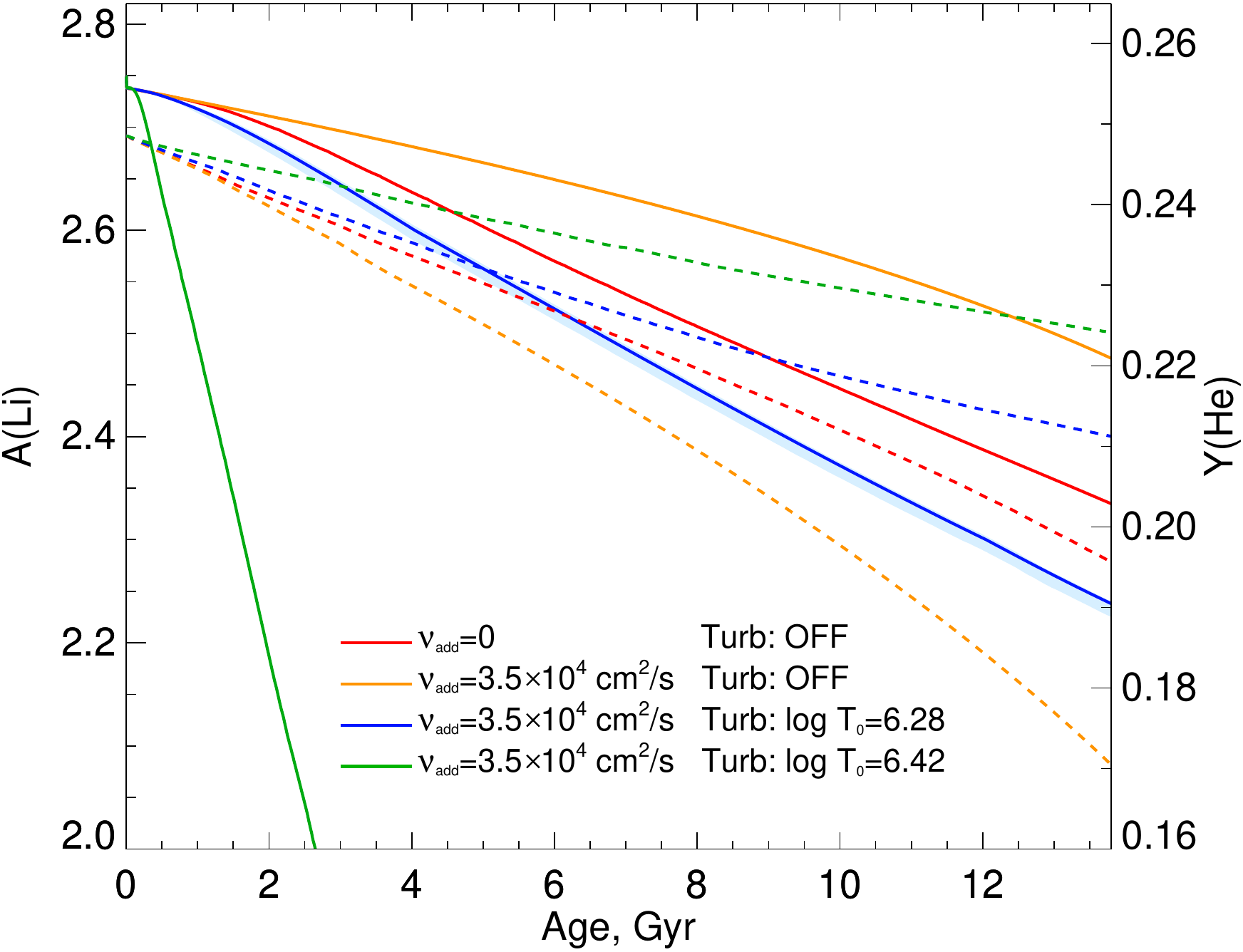}
\caption{Evolution of the surface Li abundance and $^4$He in mass fraction (solid and dashed lines respectively) as a function of age in models with M=0.74\ms, \feh=-2.0, \afe=0.3 with no additional viscosity and parametric turbulence(red line), with additional viscosity but without parametric turbulence (orange line), and with both taken into account (with log~$T_0$=6.28, blue line). The light blue band around the solid blue line indicates the A(Li) evolution for models with $\nu_{add}=(3.5\pm1.5)\times10^4$ cm$^2$s$^{-1}$ and parametric turbulence (log~$T_0$=6.28). The green lines are for a model with $\nu_{add}=3.5\times10^4$ cm$^2$s$^{-1}$ and parametric turbulence as in Pop~I stars (log~$T_0$=6.42). All the models include rotation at the median rate.}
\label{calib_models}
\end{figure}

We show in Fig.~\ref{calib_models} the impact of the $\nu_{add}$ parameter on the evolution of the Li and He surface abundances for the same reference star models. In these two specific cases, we turned off the parametric turbulence for chemicals ($D_\mathrm{T}(r)=0$, Eq.~\ref{eq_dhe}). Li depletion at the surface is due to the combined effect of atomic diffusion, meridional circulation, penetrative convection, and shear-induced turbulence, which eventually transports this fragile element from the convective envelope to slightly deeper radiation layers where it can burn. He abundance variations, on the other hand, result from the relative efficiency of the hydrodynamic processes that slow down the effects of atomic diffusion. As expected, Li is more depleted when $\nu_{add}$=0 because of the steep rotation gradients that build inside the star (Fig.~\ref{angular_velocity_profiles}), leading to more efficient shear-induced mixing, hence more Li destruction, but also less efficient atomic diffusion on He which decreases more slowly (see description of Fig.~\ref{diff_prof_rotation} below). When efficient angular momentum transport is simulated by the additional viscosity, the radial profile of angular velocity is much flatter, hence shear-induced mixing is less efficient. Li is less efficiently driven to the burning region, leading to slower surface Li depletion. On the other hand, He decreases more rapidly at the surface of the star because atomic diffusion is less counteracted by the hydrodynamical processes. In both models, however, the Li abundance at the age of $\sim$12~Gyr is higher than the assumed plateau value of 2.3~dex. As in the case of the $_{\nu}R1^{T_0}_A$ models of Pop~I stars (D21a,b), an additional process for the transport of chemicals is required to fit the Li data, that we treat below as parametric turbulence.

\subsection{Calibration of parametric turbulence and choice of the initial rotation period}
\label{subsec:turbulence_calib}

\begin{figure*}
\centering
\includegraphics[width=1\linewidth]{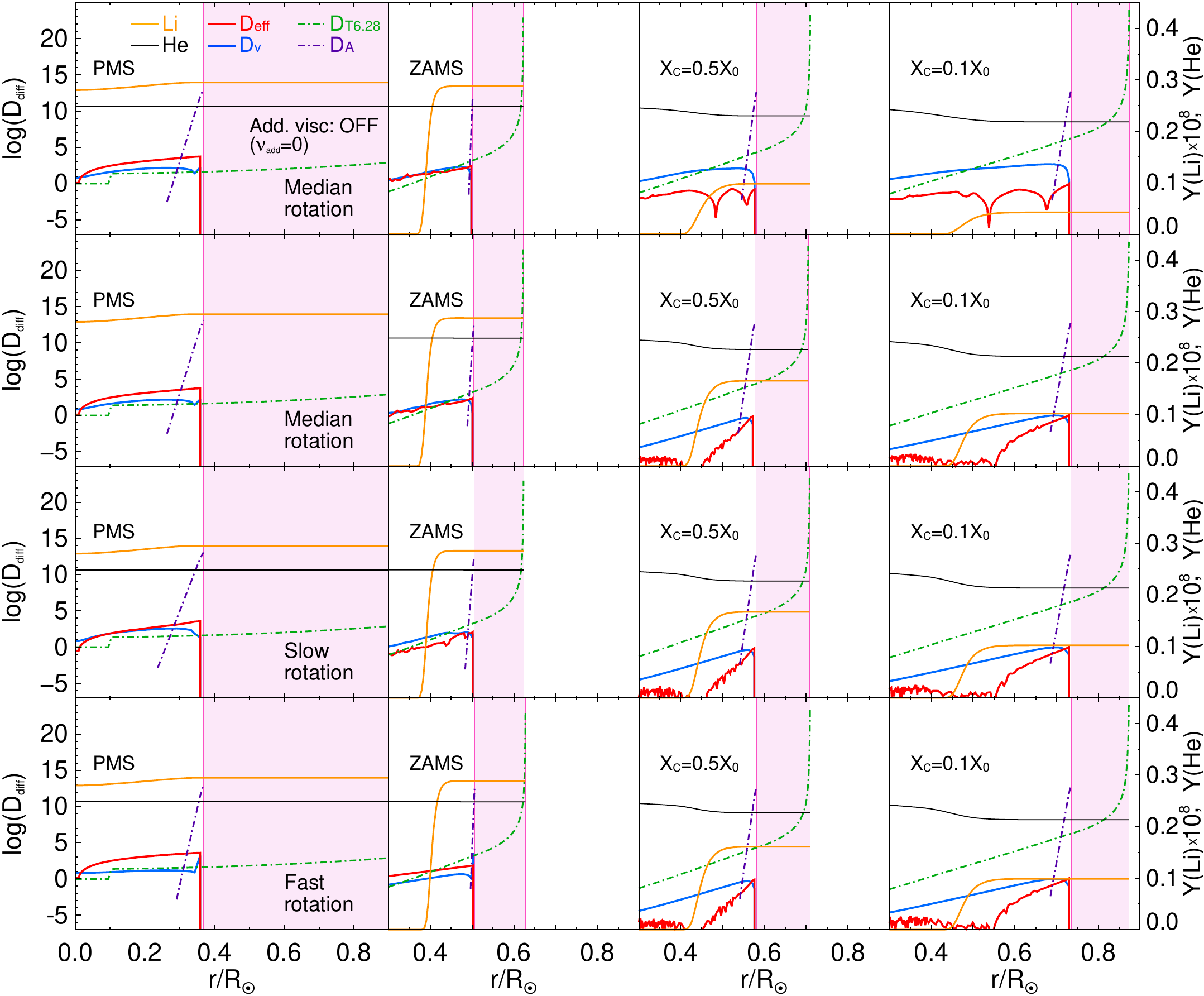}
\caption{Radial profiles of diffusion coefficients ({\it{left axis}}) of meridional circulation ($\mathrm{D_{eff}}$), vertical shear ($\mathrm{D_{v}}$), parametric turbulence ($\mathrm{D_{T6.28}}$), penetrative convection ($\mathrm{D_{A}}$) and of the Li and He mass fractions ({\it{right axis}}) for the models with M=0.74\ms \ (\feh=-2.0 and \afe=0.3) with median, slow, and fast rotation as indicated, shown on the PMS (shortly after the core became radiative, age$\approx$1.3~Myr), at the ZAMS, and on the MS when the central hydrogen mass fraction $X_C$=0.5$X_0$ and $0.1 X_0$, where $X_0$ is the initial H mass fraction (from left to right). The upper row corresponds to the model without additional parametric viscosity ($\nu_{add}$=0) and the other rows are for models computed with $\nu_{add}$=3.5 $\times 10^4$cm$^2$s$^{-1}$ and three initial rotation rates. The pinkish area corresponds to the convective envelope. For the PMS structure, we show the profiles down to the center to demonstrate the initial (almost flat) behaviour of Li in the models.}
\label{diff_prof_rotation}
\end{figure*}

As can be seen in Fig.~\ref{calib_models}, Li depletion in the reference star is too strong when assuming the $T_0$ value for parametric turbulence $D_\mathrm{T}$ (Eq.~\ref{eq_dhe}) that was calibrated in Pop~I solar-type dwarfs stars by D21b (log~$T_0$=6.42). We have adjusted the value of $T_0$ to reach the plateau value (A(Li)=2.3~dex; \citealt{Norris2023}, see \S\ref{sec:obs_data}) at the age of 12~Gyr for the reference star computed with $\nu_{add}=3.5\times10^4$ cm$^2$s$^{-1}$ and median rotation velocity. Although for calibration purposes we adopt the value of 12~Gyr as the average age of Pop~II stars, we allow an age variation of about 1~Gyr for most of them. The value that best fits our assumptions is $\log T_0$=6.28 (see discussion in \S\ref{subsec:robustness}). The time evolution of Li and He at the surface of the corresponding model is shown in Fig.~\ref{calib_models}. Parametric turbulence slows down the effect of atomic diffusion, as indicated by the behaviour of He which is less depleted at the surface than in the previous models. On the other hand, it transports more efficiently Li down to its nuclear burning temperature below the convective envelope, leading to more Li depletion than when the transport by only meridional circulation and shear turbulence were accounted for.

\begin{figure}
\centering
\includegraphics[width=1\linewidth]{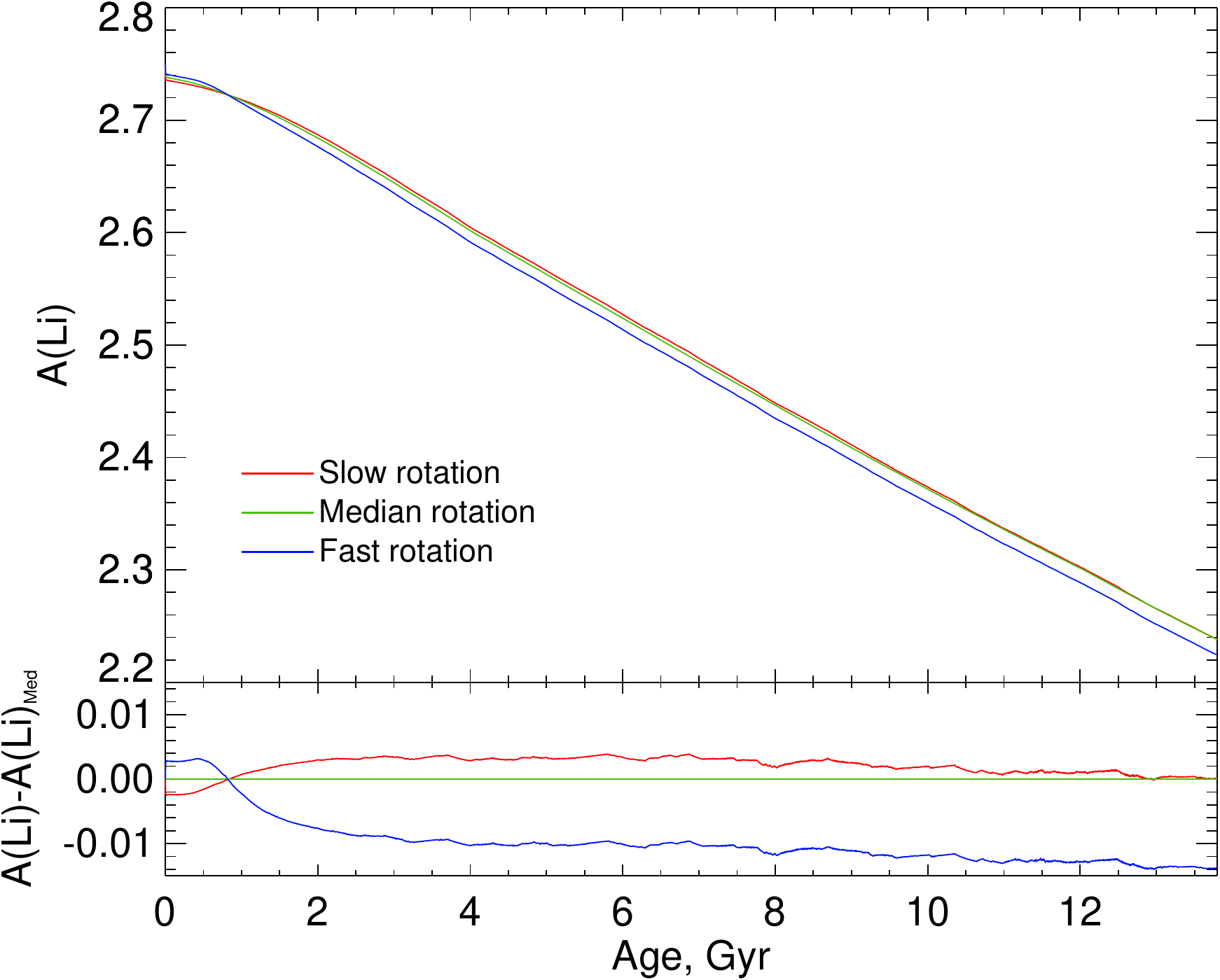}
\caption{\textit{Upper panel:} Li evolution in the $_{\nu_{add}}R1^{T_0}_A$ models with M=0.74\ms \ (\feh=-2.0 and \afe=0.3) and three different rotation rates (slow, median, and fast). \textit{Bottom panel:} Difference between Li abundance as a function of age with respect to the model with median rotation.} 
\label{diff_li_rotation}
\end{figure}

The different diffusion coefficients for chemicals are shown in Fig.~\ref{diff_prof_rotation} for the reference star models computed with different assumptions. The first three rows from the bottom are for $\nu_{add}=3.5\times10^4$ cm$^2$s$^{-1}$, parametric turbulence $D_\mathrm{T}$ with $\log T_0$=6.28, and different initial rotation rates taken in the same range as for Pop~I stars. When angular momentum extraction by the wind is efficient, i.e., on the PMS (left column, the age of each model is about 1.3~Myr) and at the beginning of the MS, the diffusion coefficients related to meridional circulation and vertical shear ($D_\mathrm{eff}$ and $D_\mathrm{v}$ respectively) are of the same order of magnitude as the diffusion coefficient $D_\mathrm{T}$.  Later along the MS, however, $D_\mathrm{eff}$ and $D_\mathrm{v}$ decrease together with the magnetic torque, and the transport of chemicals is driven by $D_\mathrm{T}$. In the middle of the MS ($X_C$=0.5$X_0$, corresponding to $\sim$7~Gyr for these models), parametric turbulence is $\sim$4 magnitudes stronger than rotation-induced mixing and becomes even stronger at later stages. While Li depletion on the PMS is slightly affected by the initial rotation rate via the depth of convective penetration through the ratio (v/v$_0$), the final Li depletion is almost independent of the initial rotation rate in the range considered for Pop~I stars (Fig.~\ref{diff_li_rotation}). At the age of 12~Gyr, the A(Li) difference between the models with slow and fast rotation is only 0.01~dex. Due to a lack of data on the initial rotation of young Pop~II stars or evidence for different behaviour, we decided to keep the median rotation rate for the grid of models. 

The first row from the top in Fig.~\ref{diff_prof_rotation} corresponds to a reference star model with $\nu_{add}=0$, parametric turbulence $D_\mathrm{T}$ with $\log T_0$=6.28, and median rotation rate. Comparison with the second row from the top illustrates the effect on $\nu_{add}$ on meridional circulation and vertical shear as discussed in \S\ref{subsec:viscosity}. In particular, the vertical shear is stronger due to the steeper differential rotation after angular momentum has been extracted by the stellar wind. Finally, we show in Fig.~\ref{calib_models} with the light blue band the Li predictions for models of the same star computed with different values of $\nu_{add}$ (2$\times 10^4$ and 5$\times 10^4$) and with the same value of the parametric turbulence calibrated to reach the plateau (compare the light blue band to the blue line). Since parametric turbulence drives the Li depletion in that case, the actual. Without internal rotation information for Pop~II stars, we thus face a slight degeneracy between the determination of the additional viscosity for the transport of angular momentum and that of the parametric turbulence for chemicals. For this reason, we decided not to explore possible variations of additional viscosity in space and time as proposed e.g. by \citet{Spada2016}.

For our grid of models, we thus assume the value of $\nu_{add}=3.5 \times 10^4$ cm$^2$.s$^{-1}$ for all our models of $_{\nu_{add}}R1^{T_0}_A$ Pop~II stars as in the R1 models of D21a,b for Pop~I stars, and median rotation with the same initial period as for Pop~I stars. The only parameter that we need to change to reach the Li plateau value is the anchoring temperature $T_0$ for parametric turbulence, with log~$T_0$=6.28 instead of 6.42 in Pop~I stars (see further discussion in \S\ref{subsec:robustness}). This is consistent with \citet{Richard2005} who determined log$T_0 \sim 6.25 - 6.28$ to best fit the constancy of the Li plateau against the effects of gravitational settling. In their study, the log~$T_0$ range came from the consideration of different temperature scales used in the observational samples they considered, which lead to different Li abundance determinations  \citep[see also][]{Charbonnel2005}, and from the mean age assumed for the sample stars. The robustness across the Li plateau (in both [Fe/H] and \teff) of the calibration we did with the  M=0.74\ms, \feh=-2.0 models is discussed below.

\begin{figure}
\centering
\includegraphics[width=1\linewidth]{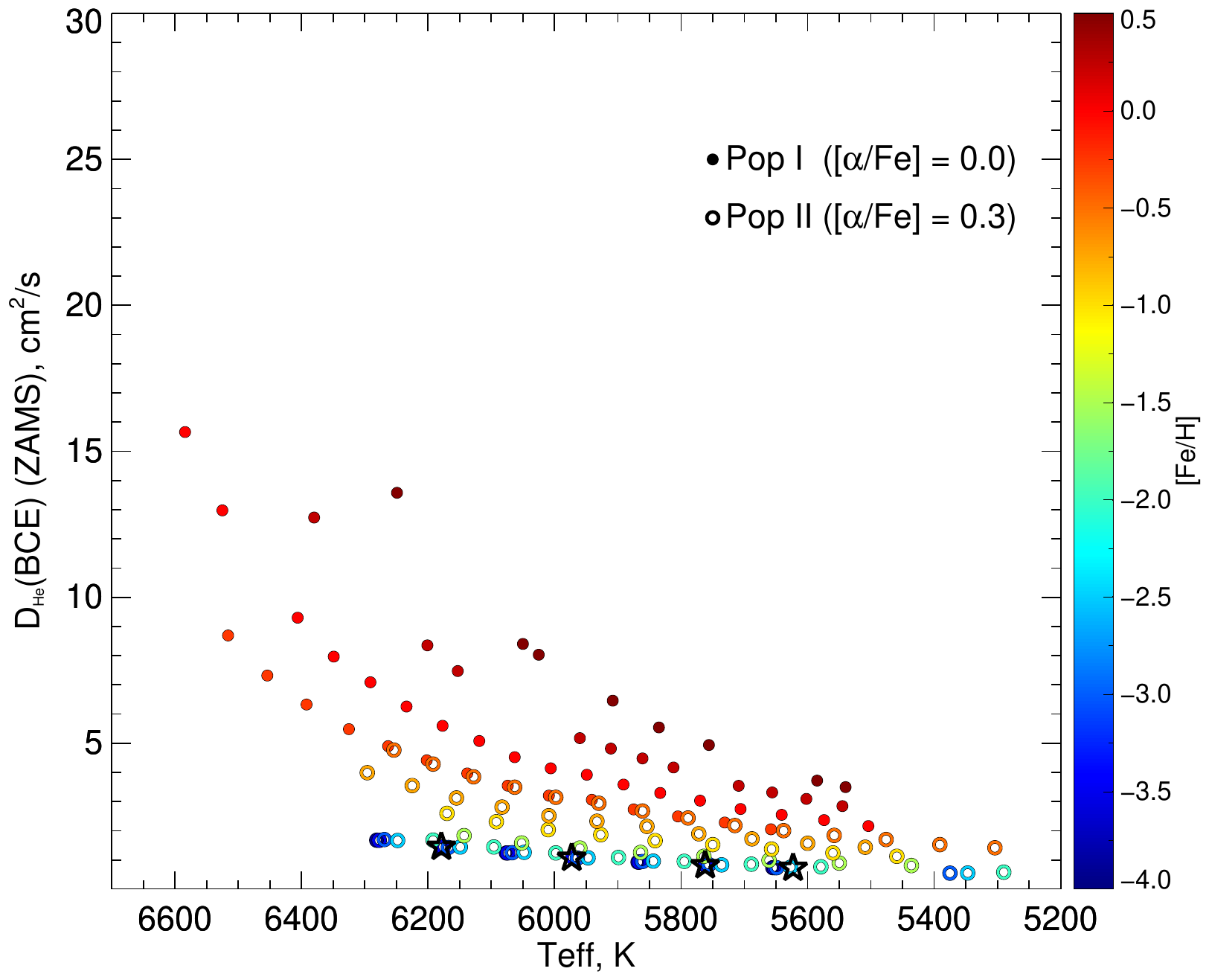}
\includegraphics[width=1\linewidth]{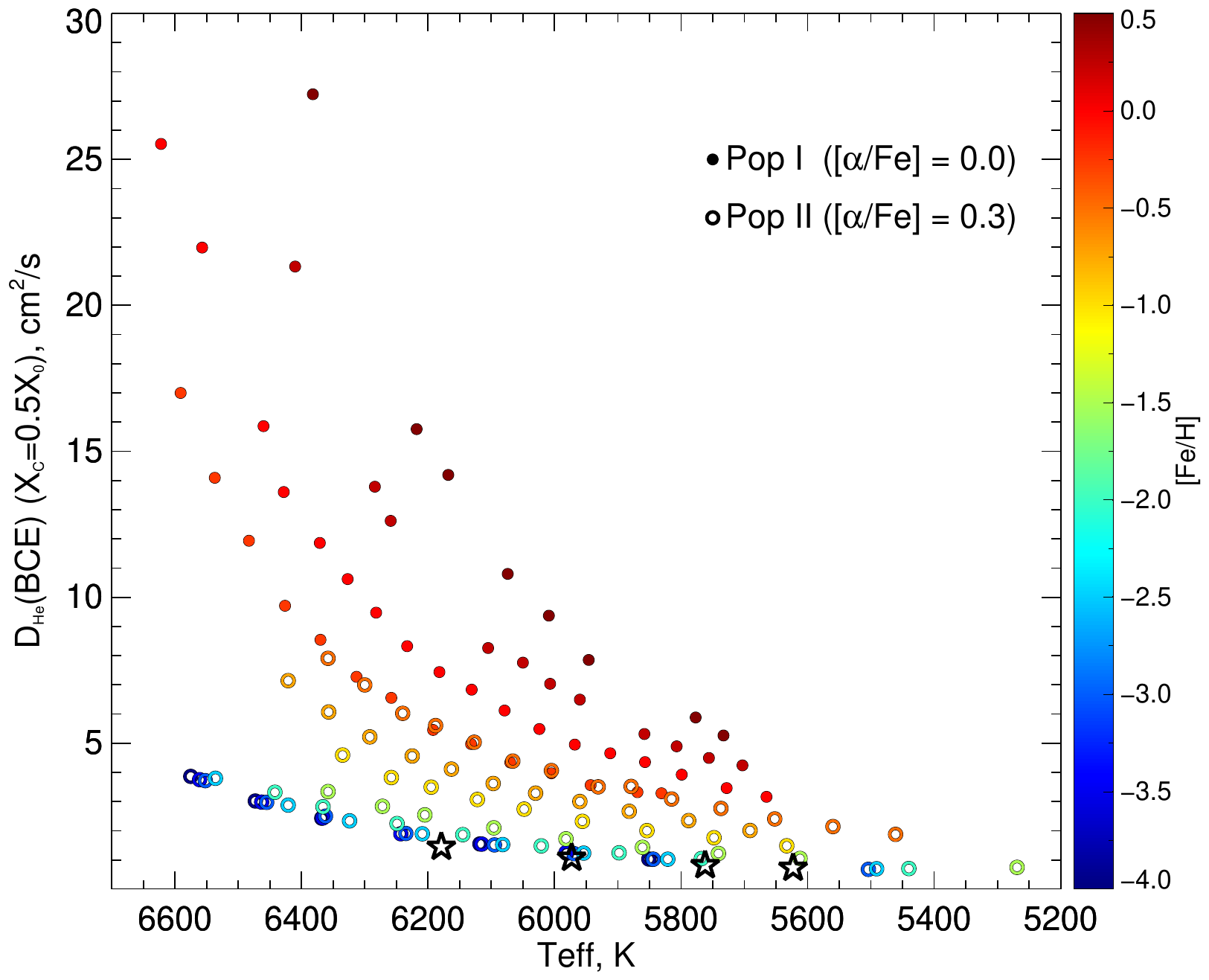}
\caption{Atomic diffusion coefficient of He just below the base of the convective envelope for Pop~II and Pop~I models (open and filled circles respectively) at the ZAMS and in the middle of the MS when $X_C=0.5X_0$ (upper and bottom panels respectively). The [Fe/H] value of the models is color-coded. With black five-pointed stars, we show the predicted values for the model with \feh=-5.8 computed for the most Fe-poor star J0023+0307 (\S~\ref{selection-field} and \ref{subsec:comp_obs_data}).}
\label{dhe_vs_teff}
\end{figure}

\subsection{Robustness of the calibration across the Spite plateau and key differences between Pop~II and Pop~I stars}
\label{subsec:robustness}

\begin{figure*}[ht]
\begin{multicols}{2}
\includegraphics[width=1\linewidth]{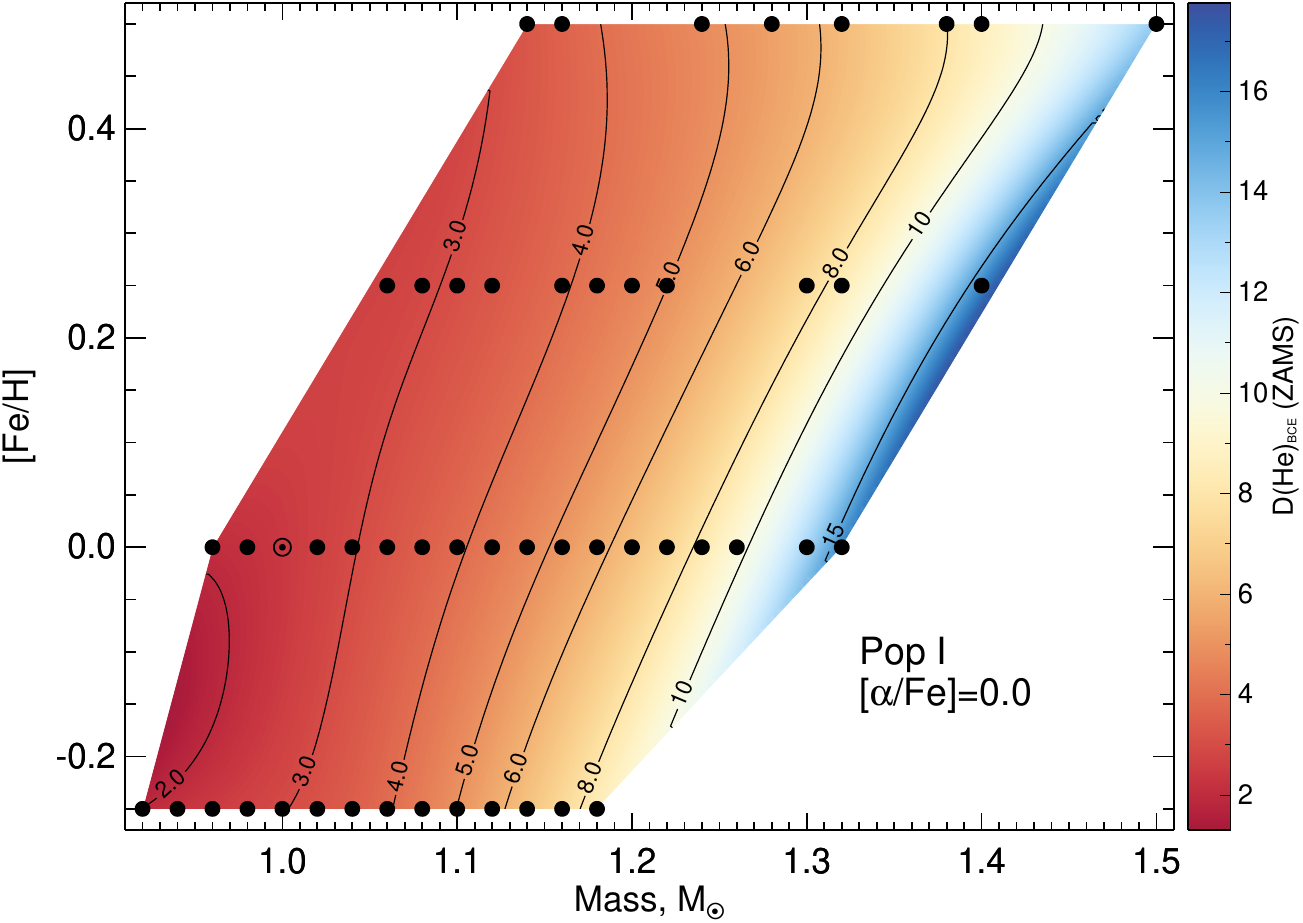}\par
\includegraphics[width=1\linewidth]{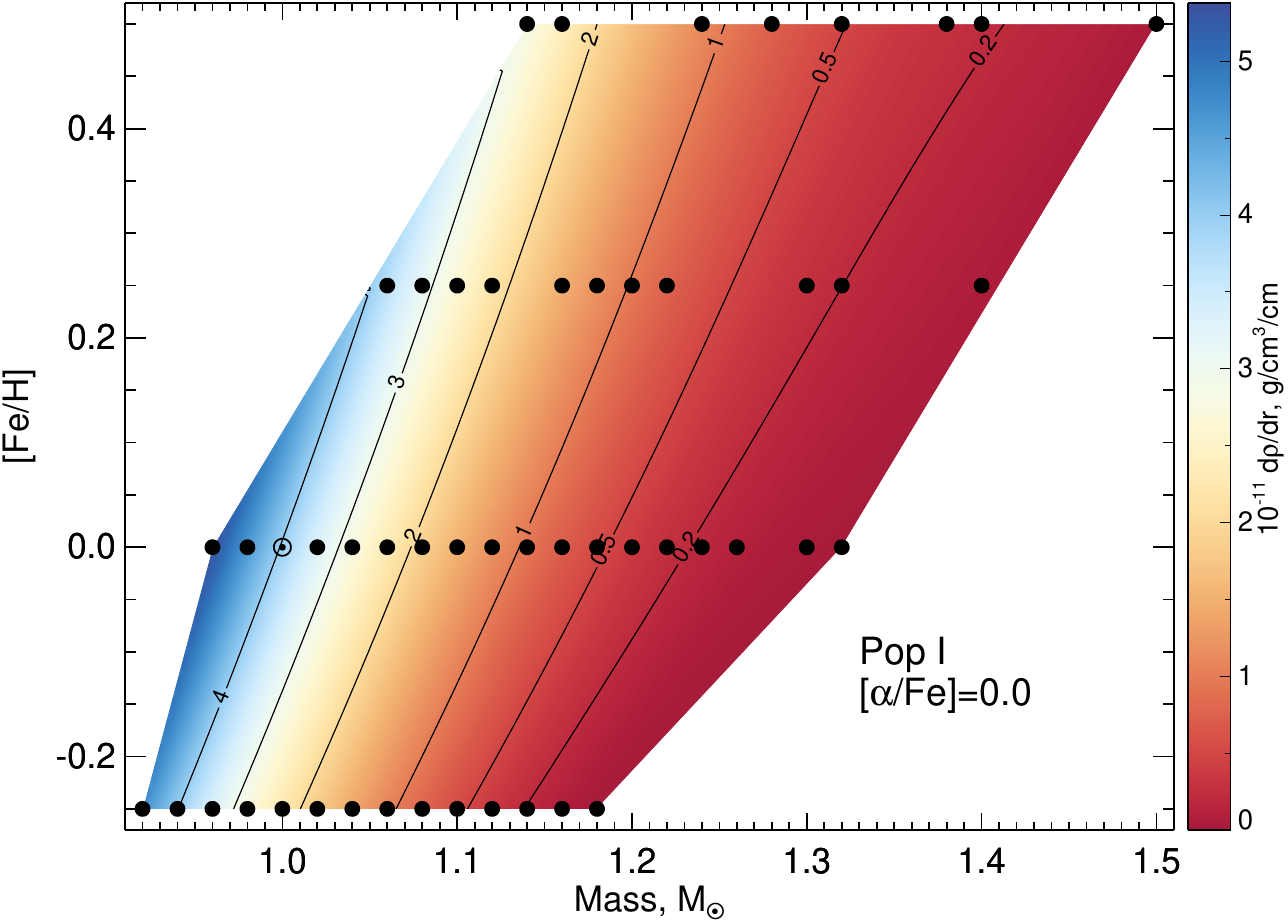}\par
\end{multicols}
\begin{multicols}{2}
\includegraphics[width=1\linewidth]{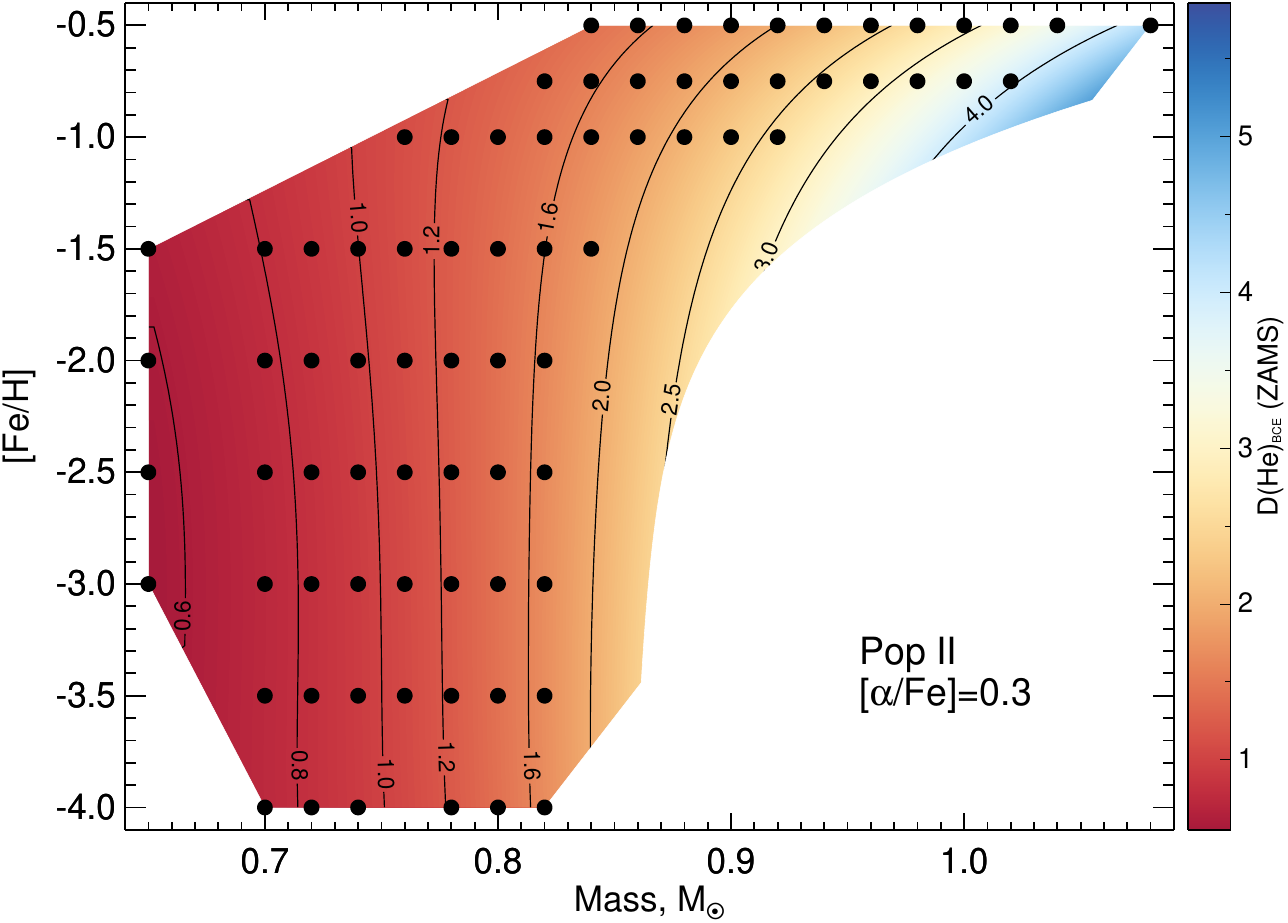}\par
\includegraphics[width=1\linewidth]{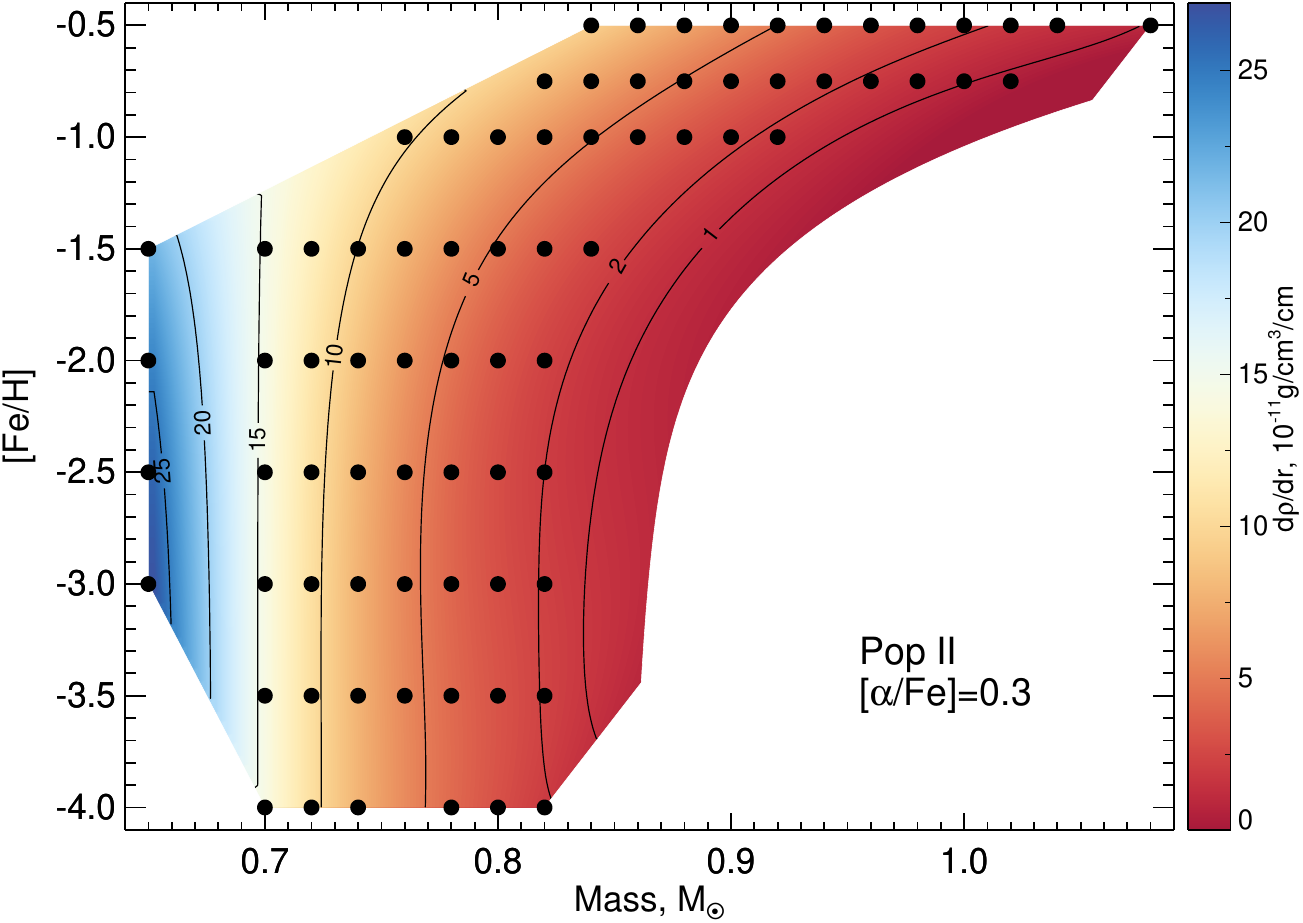}\par
\end{multicols}
\caption{Atomic diffusion coefficients of He and density gradient just below the base of the convective envelope (left and right, respectively) for the model grid of Pop~I and Pop~II stars (upper and bottom panels respectively). The circles (including the solar symbol) show the actual values for the grid points, whilst the colour-coded background shows the predicted values of the same parameters which have been interpolated with a third-degree polynomial.}
\label{grid_diffusion_zams}
\end{figure*}

In their seminal paper, \citet[][see also \citealt{2002ApJ...568..979R}]{Richard2005} discussed the properties of Pop~II stars that lead to the constancy of the Li plateau, both in \feh \ and \teff. They showed that for low-metallicity MS stars across a large range in [Fe/H], the two main quantities that drive Li settling through atomic diffusion and Li nuclear burning, namely the mass of the convective envelope and the temperature at its base, depend only on the stellar effective temperature (see Fig.2 of \citealt{2002ApJ...568..979R} and Fig.1 of \citealt{Talon2004}). With our (a priori) more sophisticated models that include rotation-induced processes but where parametric turbulence is still required to counteract atomic diffusion and transport Li to the burning region, we humbly confirm their findings and extend the comparison to Pop~I stars. We discuss in particular the behaviour of the key physical properties that potentially explain the difference in Li behaviour in the low and high metallicity regime using our entire grid of $_{\nu}R1^{T_0}_A$ Pop~II and Pop~I models.

Besides the mass and the temperature of the convective envelope, the important quantities for the combined effect of atomic diffusion and turbulence on the surface Li depletion are the density below the base of the convective envelope and the density gradient in the radiation layers down to the Li burning region. The first one determines the atomic diffusion timescale if turbulence is not present, which can be represented by the atomic diffusion coefficient just below the convective envelope. $D_\mathrm{He}$(BCE) is shown in Fig.~\ref{dhe_vs_teff} for Pop~II and Pop~I stars at the ZAMS (see also Fig.~\ref{grid_diffusion_zams}) and in the middle of the main sequence. For [Fe/H] between -4.0 and -1.5~dex, Pop~II stars are very compact, with a very similar density below the base of the convective envelope, leading to similarly small atomic diffusion coefficients at a given effective temperature and stellar mass. The models predict similar compactness down to [Fe/H]=-5.8~dex for which we have adopted the C and Na abundances of the most Fe-poor sample star (yet, this star is not a CEMP-object, see \S~\ref{selection-field} and \ref{subsec:comp_obs_data}). This explains why Li depletion for Pop~II stars of similar effective temperature is potentially independent of [Fe/H]. At the same time, density gradients $d\rho/dr$ at the BCE also show independence on metallicity up to \feh$\approx$-1.5~dex. However, they decrease together with the atomic diffusion timescale when the stellar mass increases (Fig.~\ref{grid_diffusion_zams}). These facts, combined with the prescription for parametric turbulence (Eq.~\ref{eq_dhe}), and more specifically the fact that it changes as $\rho^{-3}$, imply the existence of a relatively constant value of $T_0$ at which the equilibrium is maintained between atomic diffusion and parametric turbulence for Pop~II stars across the \teff ~and the [Fe/H] ranges of the plateau. 

When [Fe/H] increases above $\simeq$ -1.5~dex, however, the density and the atomic diffusion coefficient below the base of the convective envelope strongly vary with both [Fe/H] and \teff \ (and mass, Fig.~\ref{dhe_vs_teff} and \ref{grid_diffusion_zams}). This explains the need for a different (higher, see D21a,b) anchoring temperature $T_0$ for parametric turbulence above a certain limit in metallicity around [Fe/H]$\sim$-0.5~dex) to counteract the effect of the atomic diffusion and lead to the observed Li depletion at a given effective temperature.  

\begin{figure}
\centering
\includegraphics[width=1\linewidth]{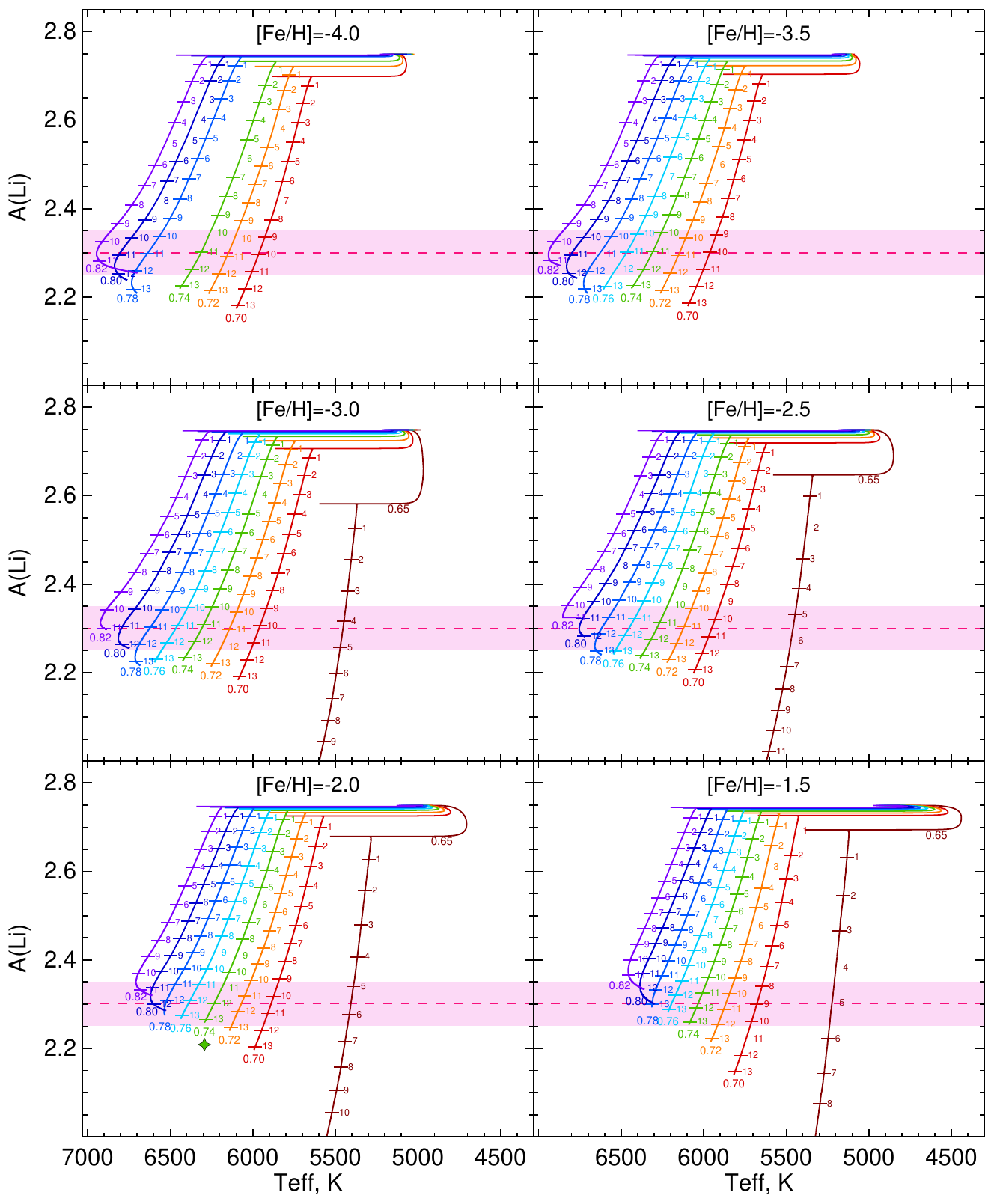}
\caption{Evolution of the surface Li abundance as a function of \teff \ for stars of different initial masses and different initial Fe content. Short horizontal ticks and numbers along the tracks indicate age in Gyr. The dashed horizontal line and surrounding pinkish area show the mean value of the Spite plateau (according to \citet{Norris2023}) and 0.05~dex scatter around it. The reference model is denoted with a four-pointed star.}
\label{li_vs_teff_poII}
\end{figure}

The evolution of the surface Li abundances as a function of \teff ~ and [Fe/H] is shown in Fig.~\ref{li_vs_teff_poII} for our grid of Pop~II $_{\nu_{add}}R1^{T_0}_A$ models. As can be seen from Fig.~\ref{calib_models}, the reference model star with \feh=-2.0 and M=0.74\ms \ reaches the value of A(Li)=2.3~dex at almost exactly 12~Gyr. At this age, the model has \teff=6234~K. While some stellar models with effective temperatures \teff$>$6000~K and an associated lithium abundance A(Li) of 2.3~dex reach this specific value at varying ages, the majority of them consistently do so within the age range of 11 to 13~Gyr. This fits within our aforementioned assumptions about age variations of Pop~II stars, and, thus, we conclude that our models can effectively describe the Spite plateau.

\section{Observational tests to model predictions}
\label{sec:obs_data}

In \S\ref{sec:calibration}, we calibrated our models assuming the plateau value to be A(Li)=2.3~dex  according to \citet{Norris2023}. We now look in more detail at their sample of carefully selected field Galactic halo stars and complement it with data for field stars from GALAH DR3 \citep{Buder2021}. We characterize the sample stars, eliminate peculiar objects, and compare to our model predictions for Li across the plateau ([Fe/H]$\leq -1.5$~dex and \teff \ $\geq 5800$~K) and in more Fe-rich dwarf stars (in what follows, we use the term ``dwarf'' for main sequence and main sequence turnoff stars). Finally, we use data in the globular cluster NGC~6752 to test our model predictions for Li and metals after the MS turnoff.

\subsection{Field dwarf stars} 
\label{subsec:obs_data_select}

\subsubsection{N23 sample}
\label{selection-field}
N23 re-investigated the shape of the Spite plateau using literature data for $\sim$200 stars with [Fe/H] between -6 and -1.5~dex, \teff \ between 6700 and 5500~K, and log~g $\gtrsim$ 3.5 (see references in N23). After a careful evaluation of the uncertainties in the Li abundances resulting from the different methods used to determine stellar \teff \ in the original papers \citep[see also][]{Charbonnel2005}, they have homogeneously re-determined the 1D,LTE Li and [Fe/H] abundances of the literature sample stars by anchoring \teff \ to the Infrared Flux Method (IRFM) of \citet{2010A&A...515L...3M}. Here we use their ``Corrected-Literature'' values as recommended by the authors, and we adopt the value of 5'800~K they find for the minimum \teff \ of the Li plateau. Also, we computed 3D,NLTE corrections with the code \texttt{Breidablik} \citep{Wang2021}. For this, we used \teff, \feh, and A(Li) values from N23 and log~$g$ computed from stellar luminosity $L$ and mass $M$ (see \S\ref{subsubsec:selection} and \S\ref{subsec:spins} for details about the computation of luminosity and mass) as $\log g = \log g_{\odot}+\log(M/M_{\odot})+4\log\left(T_\mathrm{eff}/T_\mathrm{eff,\odot}\right)-\log(L/L_{\odot})$, where we adopted $T_\mathrm{eff,\odot}$=5772~K and $\log g_{\odot}$=4.438. Although the corrections are very low (the mean absolute value is $\langle |\Delta \mathrm{NLTE}| \rangle \approx 0.016$~dex), hereafter, we will use 3D,NLTE abundance of Li for N23.

As underlined by N23, their literature sample contains stars with peculiarities potentially caused by binarity-induced processes (e.g., mass transfer from a companion) or environmental effects (e.g. very early Galactic chemical evolution under the influence of Pop~III stars, see e.g.\citealt{Piau2006}). Those should be discarded when studying the primordial Li abundance and especially when testing the stellar Li depletion solution. Following their discussion, we excluded stars classified as blue stragglers (BS) according to catalogs by \citet{Ryan2001a,Ryan2001b} and \citet{Carney2005}. We also eliminated CEMP(-no) stars from the sample by cross-checking with \citet{Lucey2023} who follow \citet{2005ARA&A..43..531B} and set the lower [C/Fe] limit for CEMP stars to +0.7~dex. Finally, we excluded binaries and chemically peculiar stars based on the classification provided by the \textsc{SIMBAD} database \citep{simbad}. Last, we applied for the remaining sample the method for dwarf selection described in \S\ref{subsubsec:selection}. As a result, the final sample we keep from N23 contains 57 ``normal'' Plateau dwarfs. 

\begin{figure}
\centering
\includegraphics[width=1\linewidth]{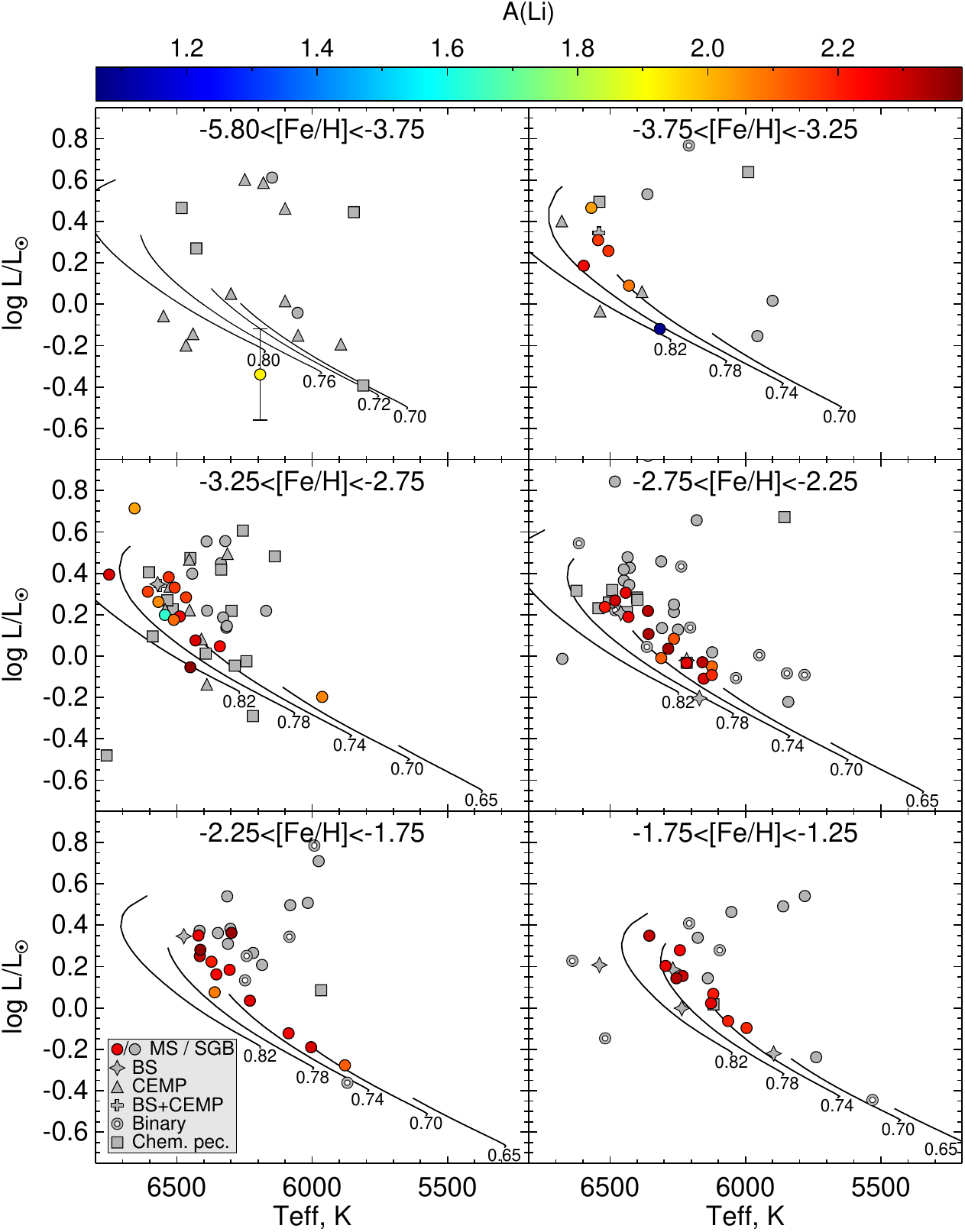}
\caption{Position in the HRD of all the N23 sample stars in different metallicity bins (we use their ``Corrected-Literature'' value for \teff ). 
Grey symbols correspond to stars on the subgiant branch (SGB) and to the peculiar objects (legend in the insert). Li abundance (3D,NLTE) is colour-coded for the 57 ``normal'' (colour-coded) Plateau dwarf stars (indicated as MS in the insert) that we use for the comparison with the theoretical Li predictions. Solid lines show the MS part of the computed evolutionary tracks computed for the middle \feh \ values of each bin except for the upper left bin that shows models at \feh=-5.8~dex that we computed assuming the C and Na content of the most Fe-poor dwarf, J0023$+$0307, for which we also show the error bars on its luminosity.}
\label{hrd_norris}
\end{figure}

\begin{figure}
\centering
\includegraphics[width=1\linewidth]{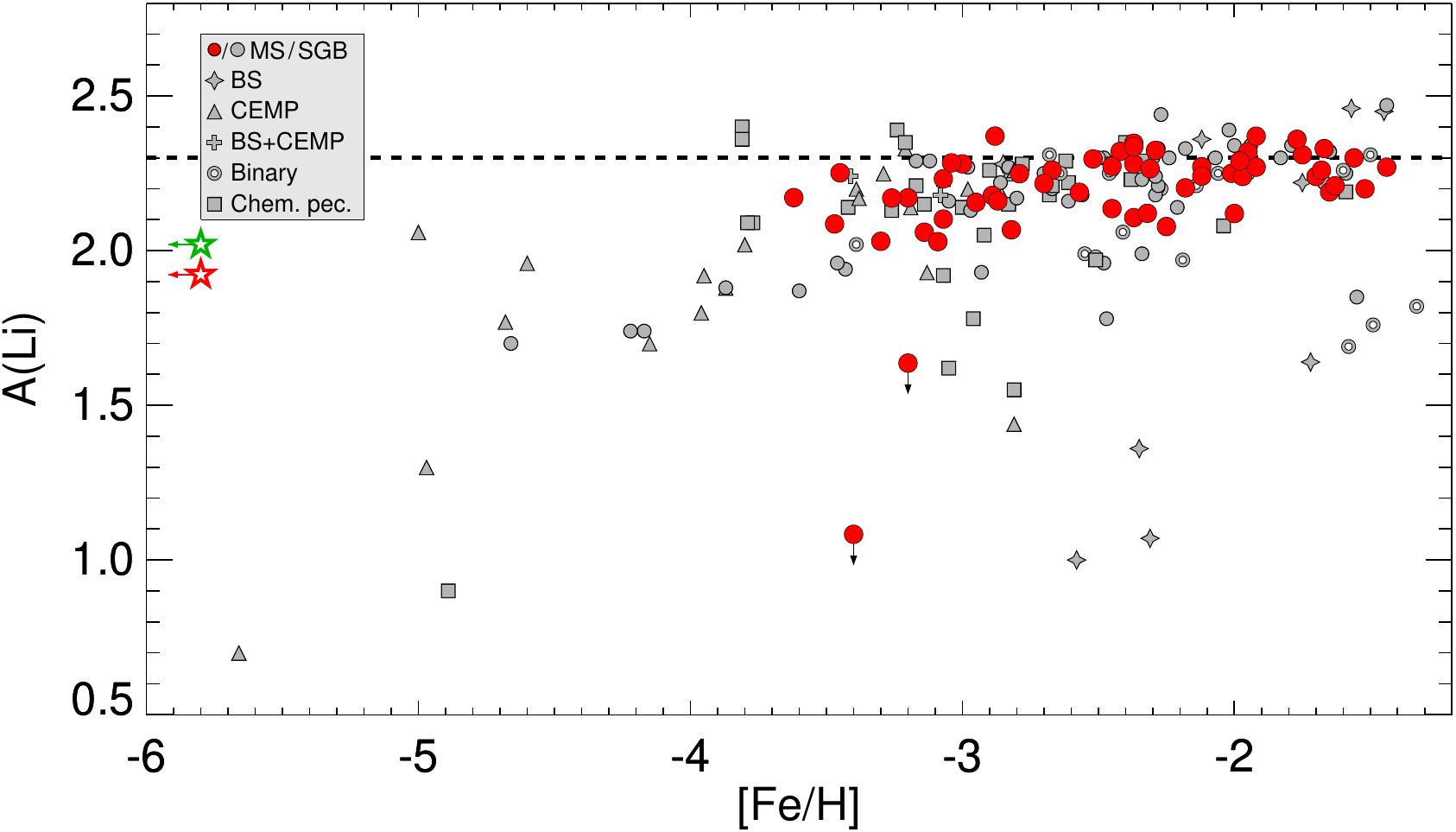}
\caption{Li abundance (3D,NLTE) as a function of \feh \ for the N23 sample (``Corrected-Literature'' values). We keep the same symbol notation as in Fig.~\ref{hrd_norris}. 
For the “normal” Plateau dwarf stars (red circles), the vertical arrows indicate Li upper limits. 
For the most Fe-poor dwarf, (shown with the red five-pointed star), J0023+0307, we also show the Li value determined by \citet[][green five-pointed star]{Aguado2019}. The horizontal arrows indicate upper limits for [Fe/H].   
The dashed line indicates the Spite plateau value of A(Li)=2.3~dex used for the calibration of our models.} 
\label{li_feh_norris}
\end{figure}

Fig.~\ref{hrd_norris} shows the position of all the stars from N23 in the HR diagram for different metallicity bins up to [Fe/H]=-1.5$\pm 0.25$~dex, with the above-mentioned peculiarities indicated by the symbol styles, and the ``normal'' dwarfs colour-coded with their Li content. Their distribution in the A(Li) versus [Fe/H] diagram is shown in Fig.~\ref{li_feh_norris}. In the [Fe/H] domain below $-$3~dex where the so-called ``meltdown'' of the Li plateau was reported \citep{Sbordone2010,Bonifacio2012,Bonifacio2015,Bonifacio2018}, the corresponding large spread of Li abundances below the plateau is due to peculiar stars that we have excluded based on the criteria described above. In particular and as underlined by N23, the incidence of C-rich stars in their original literature sample is very high in this Li-[Fe/H] domain. In the most iron-poor range ([Fe/H]$\leq$-3.75), all stars except one are extremely C-rich. These Li plateau ``outliers'' are thus expected to have been Li-deficient at birth  (see e.g. \citealt[][]{2010A&A...521A..30M,Norris2019}; N23). The most Fe-poor dwarf that belongs to this bin, J0023$+$0307, is, however, not a CEMP, and it has a Li value slightly below the plateau. The tracks we show in the corresponding panel are for tailor-made models that we computed with [Fe/H]=-5.8~dex  and the observed [Na/Fe] and [C/Fe] values (1.9 and 3.5~dex respectively, hence not a CEMP star; \citealt{Frebel2019}). This star presents abundances of $\alpha$-elements (Mg, Si, Ca, Ti) at very different levels (from [Ca/Fe]$\sim$0~dex to [Mg/Fe]$\sim$2.8~dex). For this reason, we computed models with three different values of $\alpha$-enhancement -- \afe=0.5, 1.0, and 2.0~dex. However, the difference between models with different \afe \ is negligible, and we show only models with \afe=1.0~dex. In our following discussion about the stellar depletion solution, we keep only the 57 ``normal'' (colour-coded) Plateau dwarf stars (with \teff \ $\geq 5800$~K, and [Fe/H]$\leq$-1.5~dex).

\subsubsection{GALAH sample}
\label{selection-fieldGALAH}

We selected Plateau field dwarf stars from the GALAH~DR3 sample with -2.84$\leq$\feh$\leq$-1.25~dex and  \teff \ $\geq 5800$~K. The lower limit of the metallicity range corresponds to the most metal-poor star in GALAH~DR3 after applying all the selection criteria (see below). 
We applied the quality flags on stellar parameters (\teff \ and log~g), Li abundance and \feh: \texttt{flag\_sp}=0, \texttt{flag\_li\_fe}=0, \texttt{flag\_fe\_h}=0. The \texttt{flag\_sp} criterion is very likely to eliminate peculiar stars as those we excluded in N23 sample. Also, we selected only objects with S/N per pixel $\geq$30 as recommended by \citet{Buder2021} and age uncertainty $\sigma_{age}$/age$\leq$10\% with age (and $\sigma_{age}$) from the value-added catalog of GALAH~DR3. We kept only stars with low values of interstellar extinction ($A_G\leq0.2$). As was discussed in \citet{Charbonnel2021}, this parameter might be in a strong degeneracy with \teff, which affects the precision of the latter and luminosity, thus causing an incorrect determination of age. Finally, to select dwarfs, we initially selected stars with log~$g\geq3.8$ and then performed the dwarf selection algorithm to eliminate possible contamination by subgiants (see \S\ref{subsubsec:selection}). The final GALAH sample of dwarf stars along the [Fe/H] and \teff \ domain of the Li plateau consists of 72 stars. It is worth mentioning that, unlike the N23 sample, GALAH~DR3 provides 1D,NLTE abundances of Li. 

GALAH DR3 contains a large number of dwarf stars at higher metallicities, including the metallicities of Pop~I stars even at super-solar metallicities, which is necessary for the comparison of our model predictions with observational data in the high-metallicity regime (see \S\ref{subsec:comp_obs_data_highZ}). For this purpose, we completed the sample with stars at \feh$>$-1.25~dex and log~$g\geq4.0$ that meet the above mentioned quality flags. The additional ``high-metallicity'' subsample covers the \teff \ range of the plateau from 5800~K to 6500~K. 

\subsubsection{Final step of the selection of the dwarfs}
\label{subsubsec:selection}

\begin{figure}
\centering
\includegraphics[width=1\linewidth]{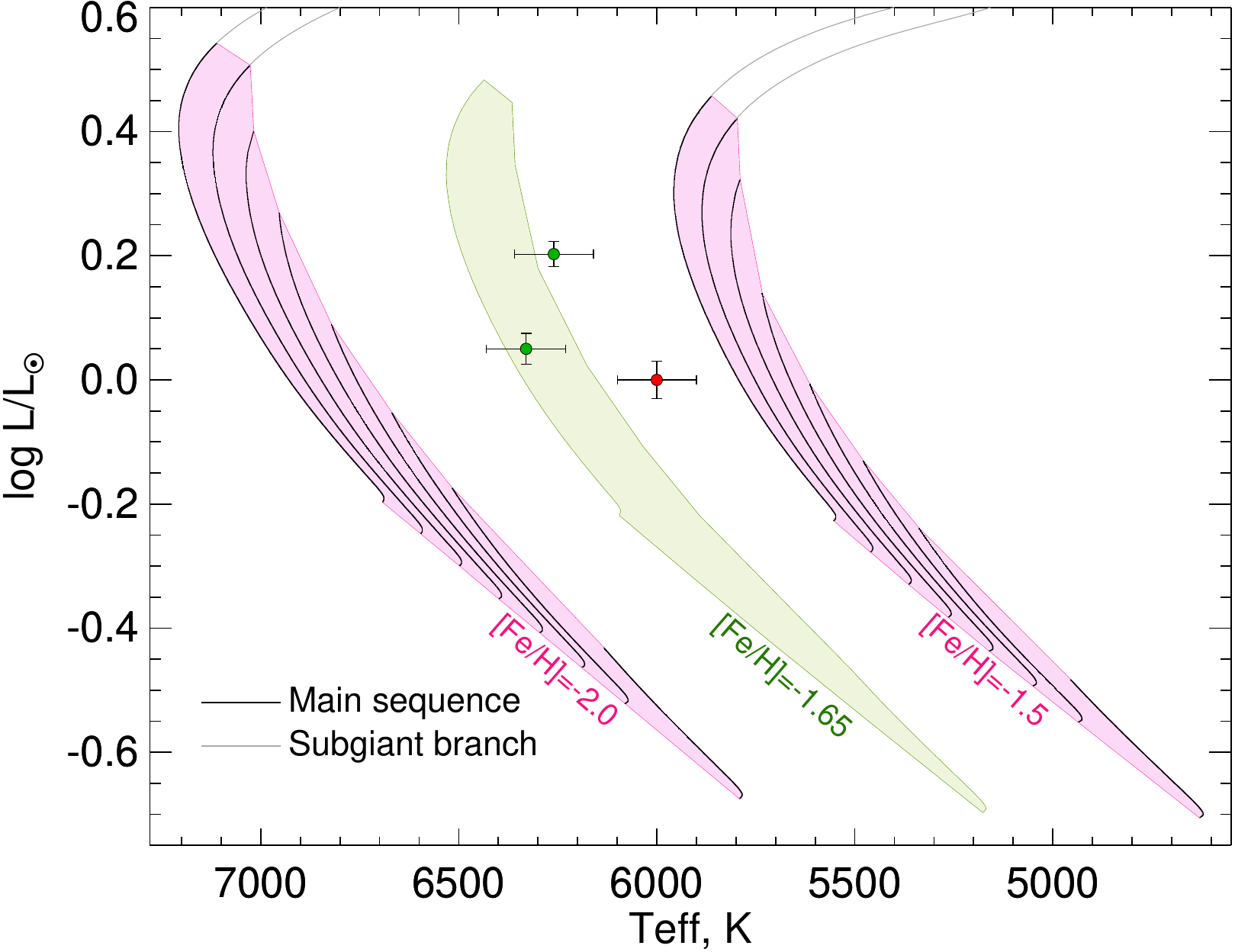}
\caption{Process of selection of dwarfs. Lines show evolutionary tracks for masses between 0.65 and 0.82\ms \ and metallicities of [Fe/H]=-2.0 (left group tracks) and [Fe/H]=-1.5 (right group of tracks). For better visibility, the tracks for [Fe/H]=-2.0 and Fe/H]=-1.5 were shifted by 500~K to the left and the right correspondingly. Black parts of the lines and pinkish areas indicate the main sequence, while grey lines show the subgiant stage. The green area is an interpolation of the main sequences from the presented tracks to the metallicity of [Fe/H]=-1.65. The points show mock stars that have different positions with respect to the interpolated main sequence area: one star is completely outside of the area even with its uncertainties,  another one is outside but hits the area with its \teff \ uncertainty, and the last one is inside the area (last two are shown in green).}
\label{select_ms}
\end{figure}

For both the  N23 and GALAH samples, we performed the following procedure to carefully select dwarf stars (MS and MS turnoff). First, we computed the stellar luminosities of the sample stars using the Gaia~DR3 data. For this, we took into account bolometric correction $BC_G$ computed with the tool by \citet{Creevey2023} and interstellar extinction $A_G$: $L/L_{\odot}=10^{0.4\cdot(M_{\odot}^{bol}-(G+5-5\log(d)+BC_G-A_G))}$, where $M_{\odot}^{bol}=4.74^m$ \citep{prsa2016}, $G$ is the apparent magnitude, and $d$ is the geometric distance \citep{bailer-jones2020}. Then, with the use of our evolutionary tracks, we defined the main sequence (MS) area in the Hertzsprung-Russell Diagram (HRD) for each metallicity of the grid (see Fig.~\ref{select_ms}). This area is represented by a polygon with Zero-Age Main Sequence (ZAMS) as vertices on one side and either the Terminal Age Main Sequence (TAMS) for stars more massive than $\sim$0.8\ms \ or points with the age of 13.8~Gyrs (age of the Universe according to \citealt{Planck2015}) for the less massive ones. In STAREVOL, the TAMS is defined as a moment when central hydrogen content X$_C$ drops to $10^{-7}$. The last step is to determine whether an individual star is inside or outside the MS area. However, for consistency, for each star, we performed linear interpolation of the position of all the vertices on the corresponding value of \feh \ and \afe \ of the star. We considered that a star is a dwarf if the values of its luminosity and \teff \ fall into the polygon at least within the 1-$\sigma$ confidence interval. However, if the polygon and the star's luminosity and \teff \ confidence intervals do not overlap, we marked this star as non-dwarf.

\begin{figure}
\centering
\includegraphics[width=1\linewidth]{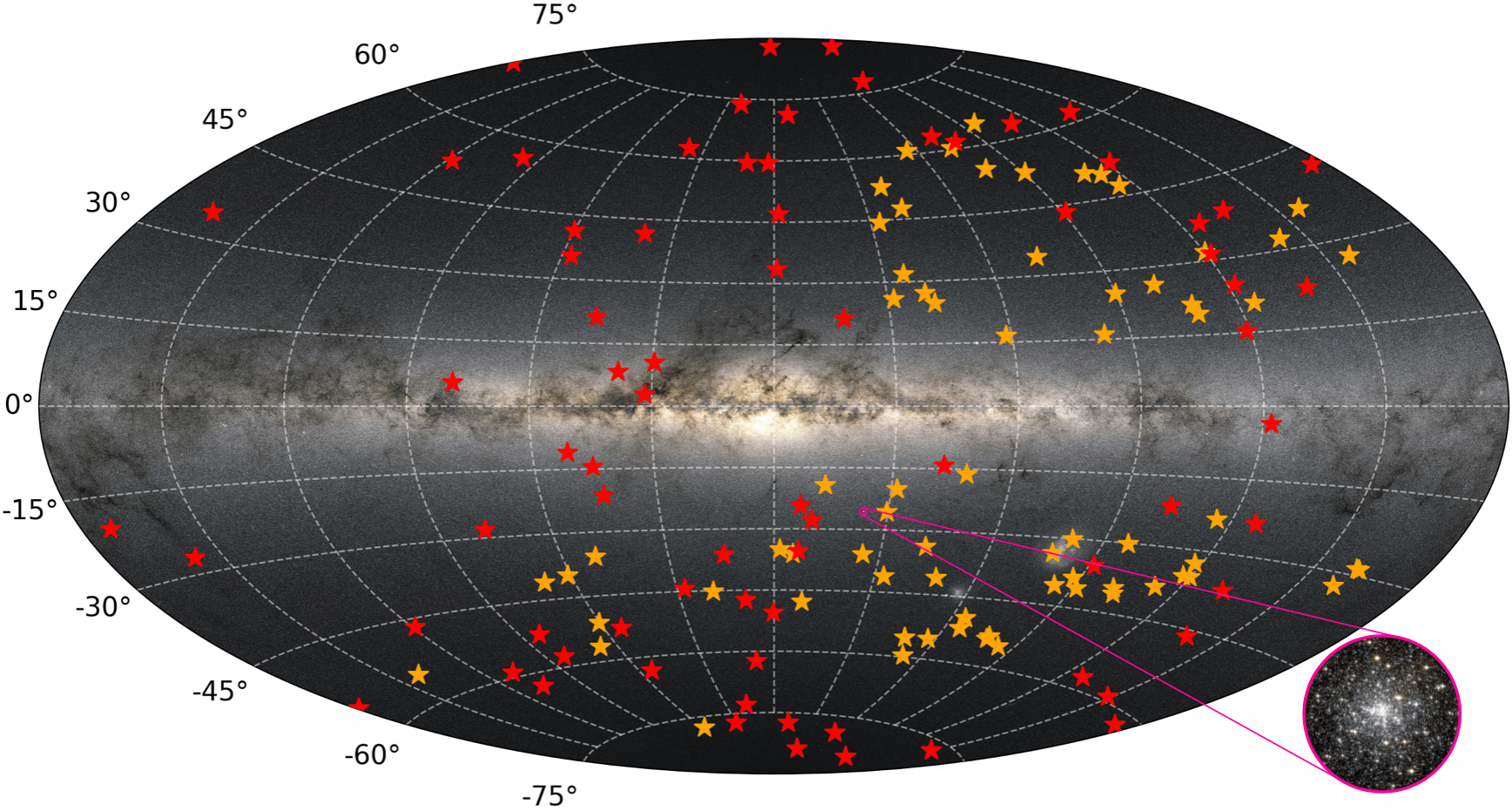}
\caption{Milky Way map with the Pop~II dwarf stars shown as red and orange points (the N23 and GALAH subsamples respectively). The inset shows the globular cluster NGC~6752.}
\label{mw_map}
\end{figure}

\begin{figure}
\centering
\begin{multicols}{2}
\includegraphics[width=\hsize]{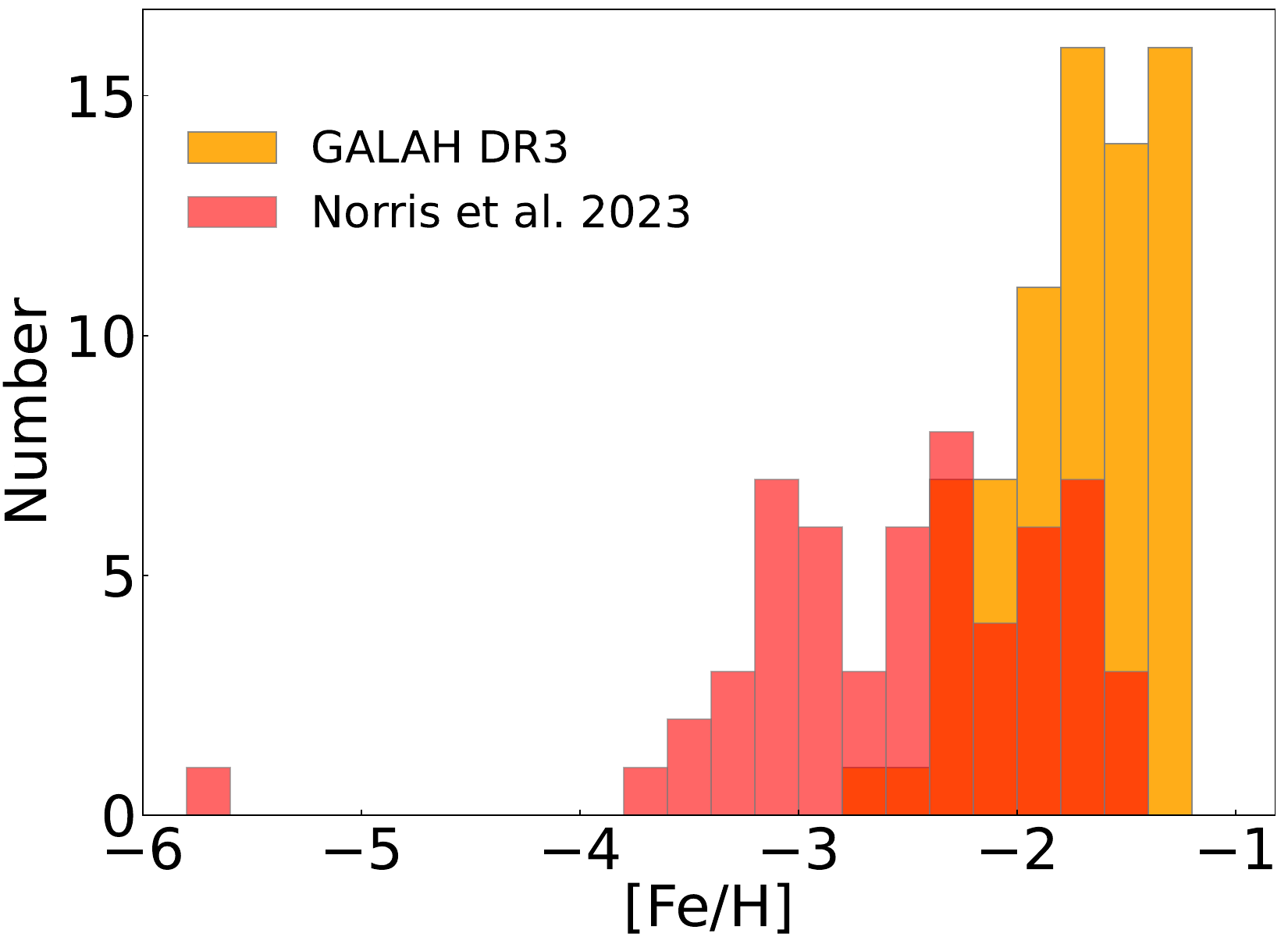}\par
\includegraphics[width=\hsize]{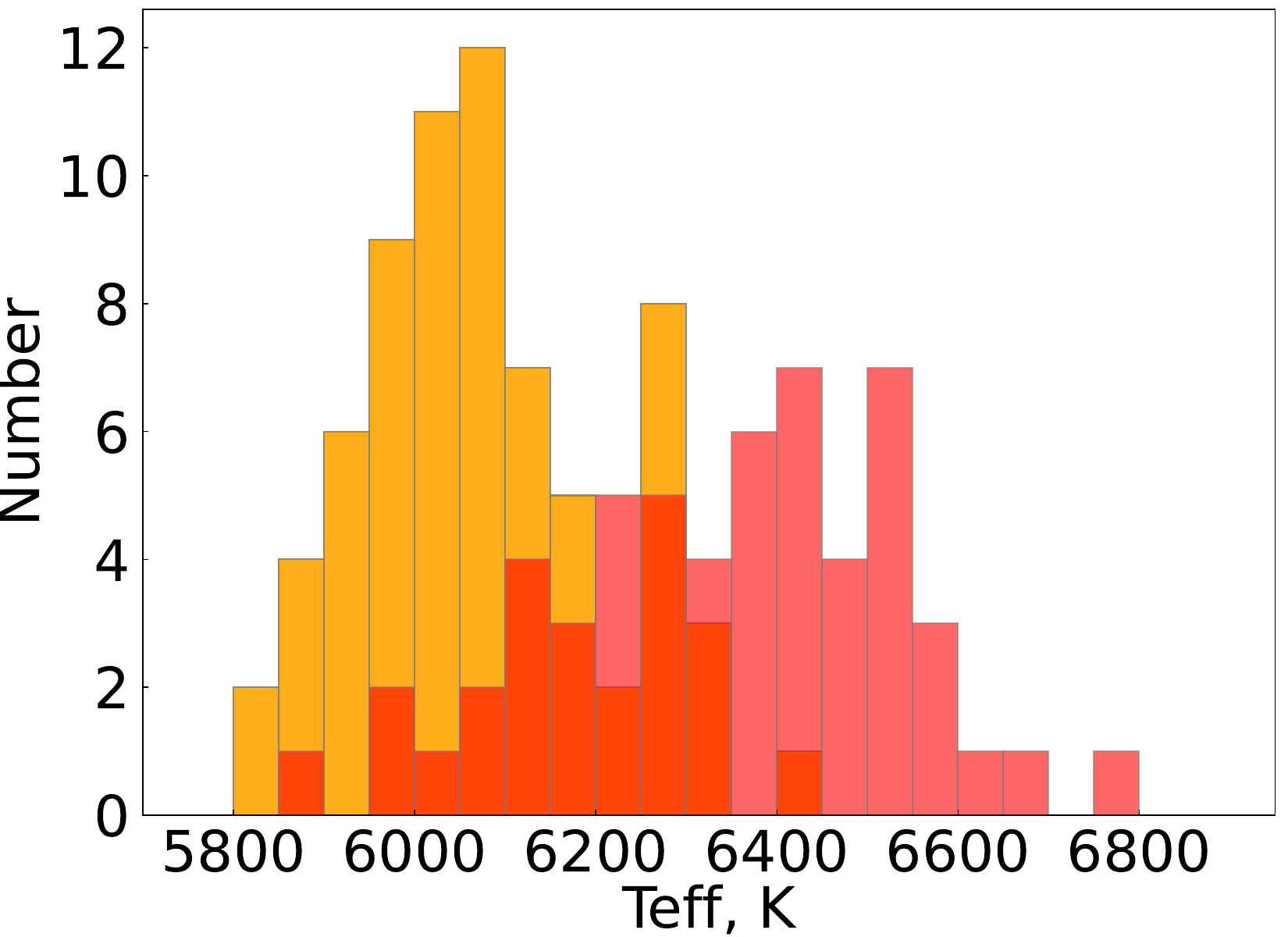}\par
\end{multicols}
\begin{multicols}{2}
\includegraphics[width=\hsize]{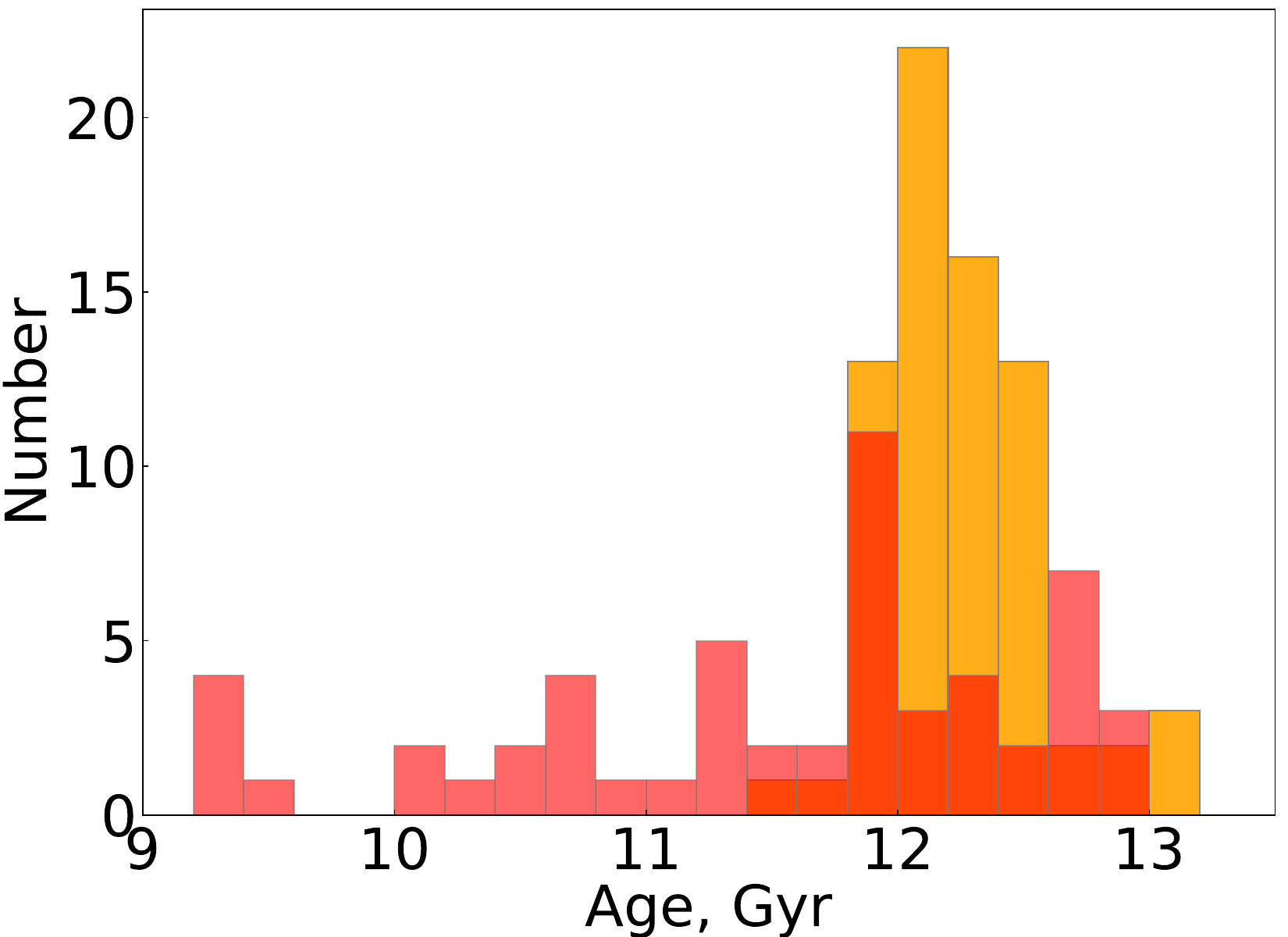}\par
\includegraphics[width=\hsize]{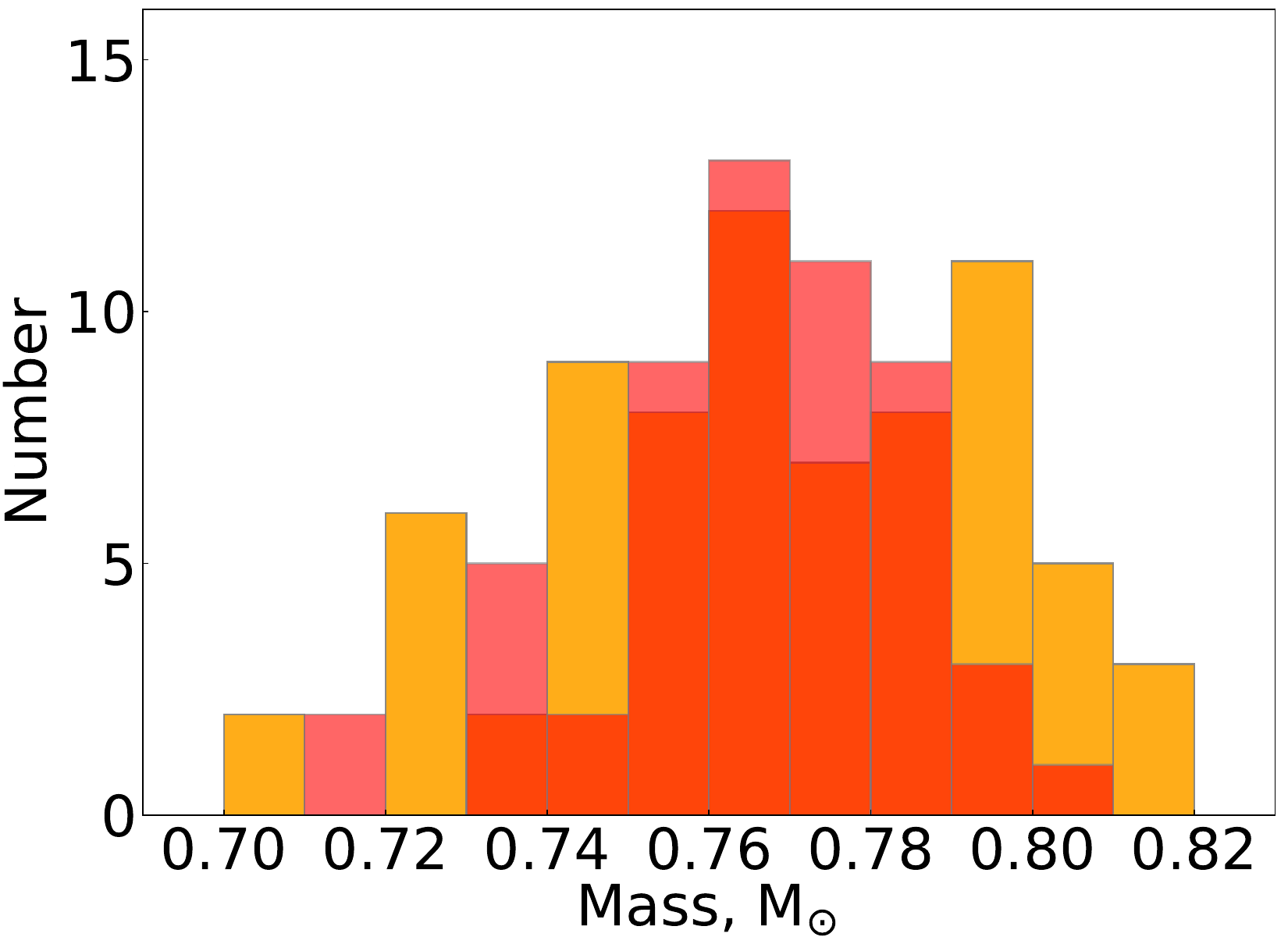}\par
\end{multicols}
\caption{\feh, \teff, age, and mass distributions of the ``normal'' dwarf stars in the [Fe/H] and \teff \ ranges of the plateau from the GALAH and N23 samples (orange and red histograms respectively).}
\label{distrib_mass_age}
\end{figure}

The \feh \ and \teff \ distributions of both the N23 and GALAH Plateau samples are presented in the upper panels of Fig.~\ref{distrib_mass_age}. Their positions, as well as the position of the globular cluster NGC~6752 (see \S\ref{subsec:selection-GC}), in the Milky Way map are shown in Fig.~\ref{mw_map}. In this figure, it is seen that the vast majority of the selected stars are located well above the mid-plane of the Galactic disc, i.e. in the halo ($\sim$94\% of the samples stars have galactic latitude $|b|>$15$^{\circ}$).

\subsubsection{Determination of masses and ages of the sample stars}
\label{subsec:spins}
We computed the masses and ages of all the field dwarf stars we selected from N23+GALAH (except for J0023$+$0307) using SPInS \citep{Lebreton2020} with our present grid of models, following the steps described in \citet{Borisov2022}. For luminosities, we used the values that we previously computed and described in \S\ref{subsubsec:selection}. For the stars from N23, we adopted values of 100~K, 0.1~dex, and 0.2~dex for uncertainties of \teff, \feh, and log~$g$ respectively. For the GALAH sample, we used the uncertainties from the original publication for individual stars. 

\begin{figure}
\centering
\includegraphics[width=1\linewidth]{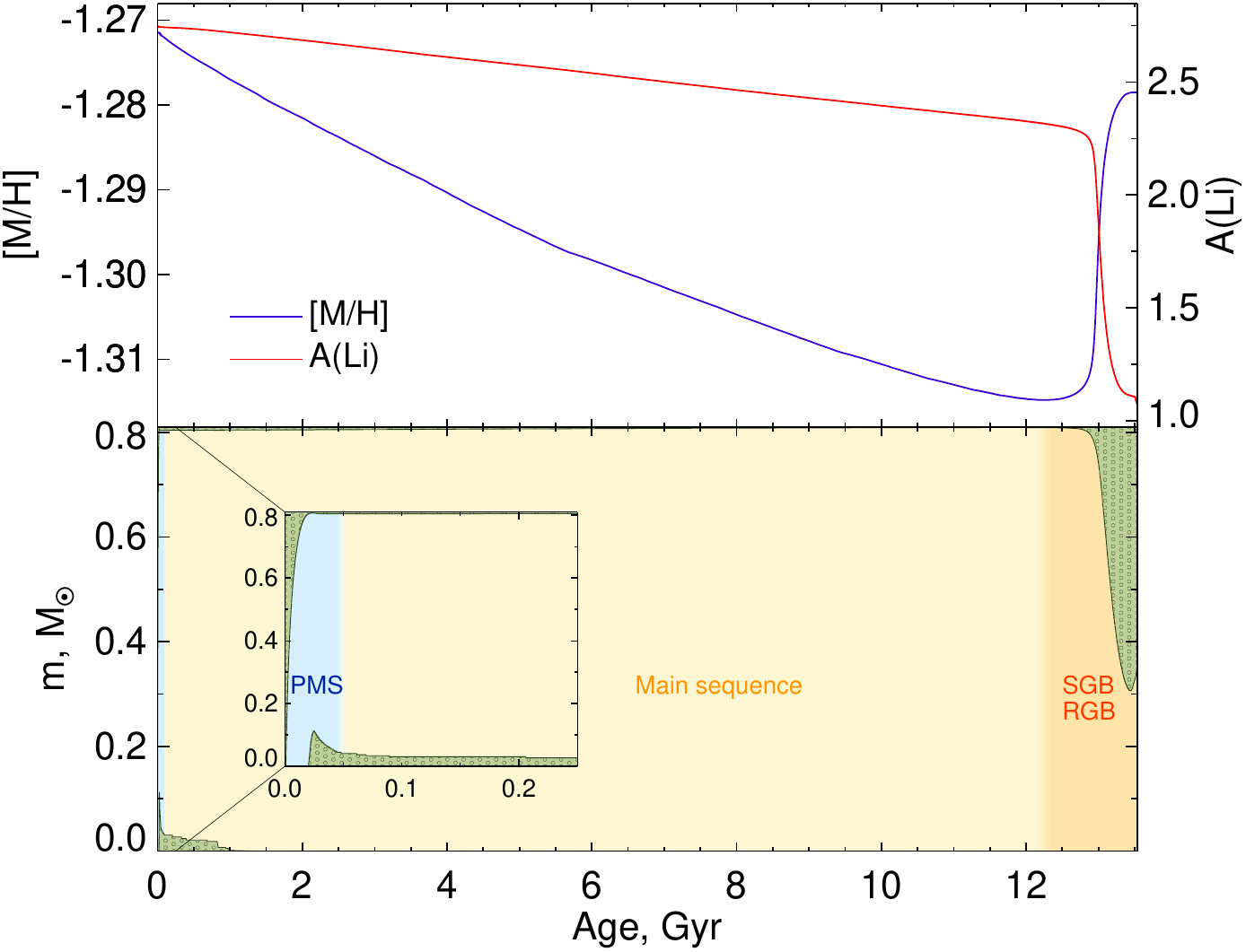}
\caption{\textit{Upper panel:} Evolution of metallicity [M/H]=$\log(Z/Z_{\odot})$ (blue line, left axis) and Li content A(Li) (red line, right axis) in the M=0.81\ms \ star with [Fe/H]=-1.5 and \afe=0.3. \textit{Bottom panel:} Kippenhahn diagram for the same model star. The green areas indicate the convective regions. Light blue, yellow, and orange areas show the radiative zone at the pre-main sequence, main sequence, and subgiant stages correspondingly.}
\label{moh_li_kippen}
\end{figure}

We accounted for metallicity [M/H]=$\log(Z/Z_{\odot})$ variations during the evolution due to atomic diffusion and other transport processes. Although this effect does not significantly change the surface metallicity of a star in the domain we explore  ($\Delta$[M/H] rarely exceeds 0.05~dex), it is known to decrease the theoretical ages and increase the estimated masses compared to models without atomic diffusion \citep[e.g.][]{2002ApJ...571..487V}. Fig.~\ref{moh_li_kippen} shows the evolution of the surface [M/H] and A(Li) and of the structure of a star with M=0.81\ms, \feh=-1.5, and \afe=0.3. It is clearly seen that once the star leaves the MS and becomes a subgiant, its convective envelope rapidly deepens (so-called first dredge-up) and brings back to the surface the metals that have ``sunk'' in the radiative layers along the MS. This is, however, not the case for Li as it has not accumulated below the convective envelope but has burned at relatively low burning temperature where it was transported by the turbulence (see \S\ref{subsec:turbulence_calib}).

The bottom panels of Fig.~\ref{distrib_mass_age} show the age and mass distributions of the Pop~II dwarf stars selected from the samples of N23 and GALAH in the [Fe/H] and  \teff \ regime of the Li plateau. The age of the selected sample stars is between $\sim$9 and 13~Gyr, with a peak around 12~Gyr (as the typical age we assumed in \S\ref{subsec:turbulence_calib}). Their [Fe/H] goes from around -4~dex, with only J0023$+$0307 with [Fe/H] below -4~dex (see \S\ref{selection-field} and \ref{subsec:comp_obs_data}). For this star, we did not determine its mass and age but a rough visual estimation based on its position in HRD (see Fig.~\ref{hrd_norris})suggests its mass lies approximately between 0.72 and 0.76\ms.

\subsubsection{Comparison between models and observational data in the \teff \ - [Fe/H]   regime of the plateau}
\label{subsec:comp_obs_data}

\begin{figure}
\centering
\includegraphics[width=1\linewidth]{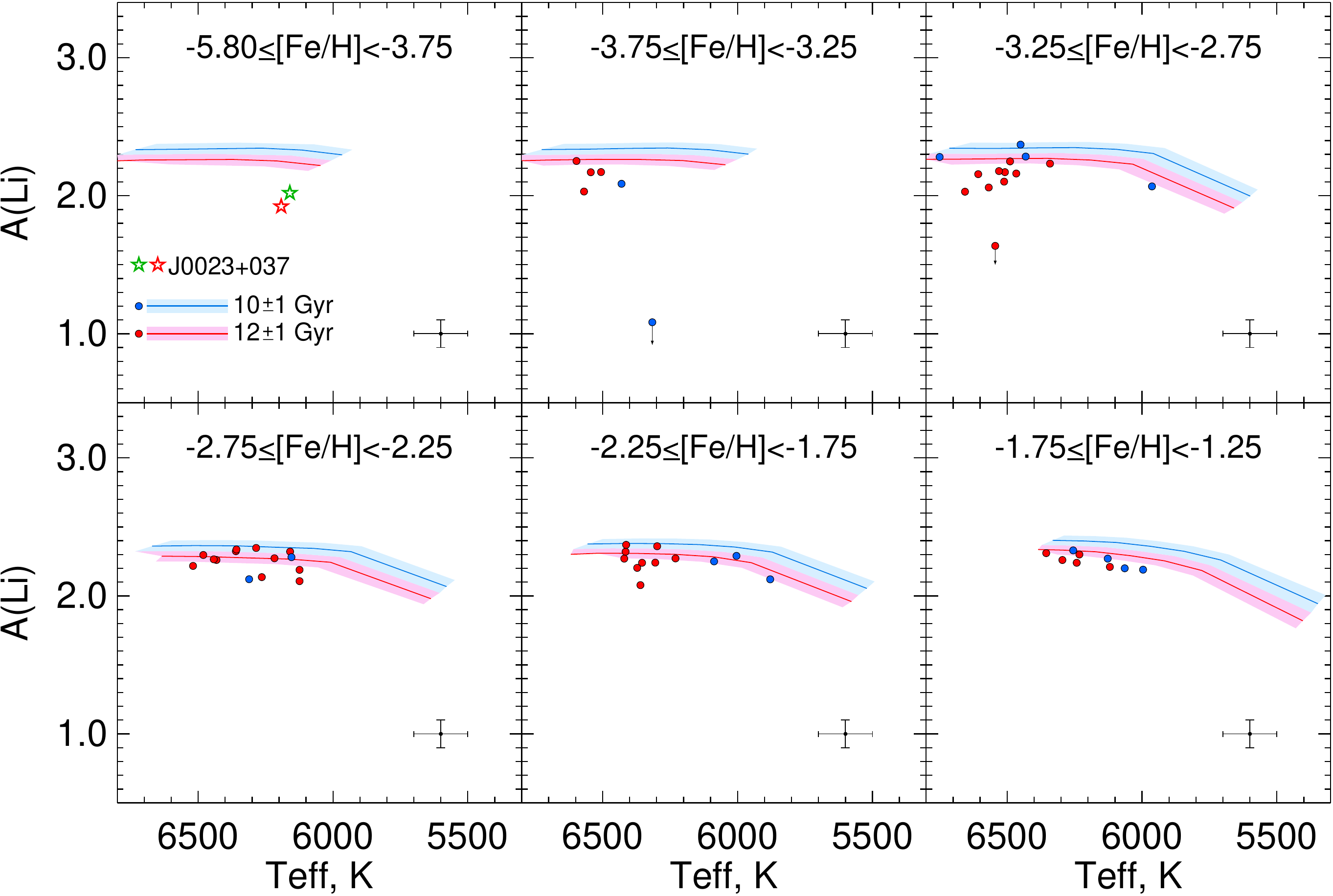}
\includegraphics[width=1\linewidth]{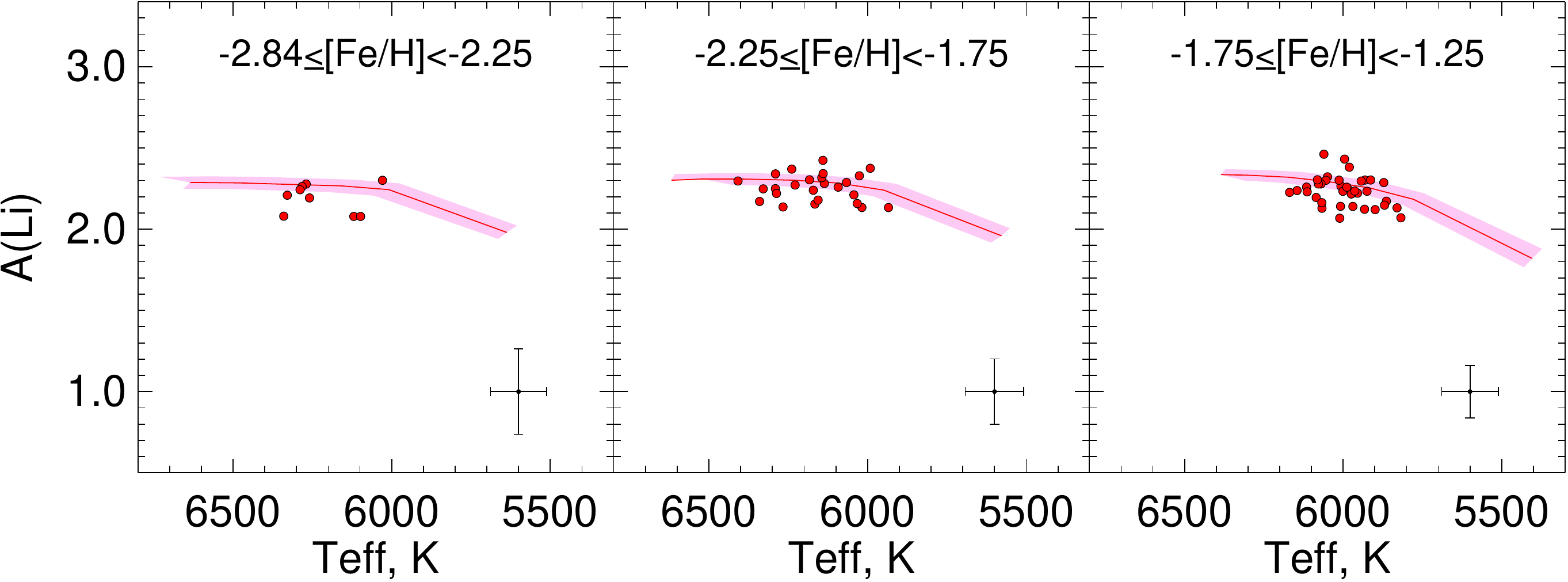}
\caption{Comparison between Li observations and theoretical Li isochrones as a function of \teff \ in different [Fe/H] bins. Blue and red isochrone lines and data points correspond to two age ranges,  10$\pm$1 and 12$\pm$1~Gyr respectively. Circles and arrows correspond to Li abundance determinations and upper limits respectively. Error bars show mean \teff \ and A(Li) uncertainty of stars in a given bin. 
\textit{Top and bottom:}  N23 and GALAH sample stars respectively. In the upper left panel, we show the position of the star most Fe-poor dwarf of our sample, J0023+037, with the Li and \teff \ values from N23 (red five-pointed star) and from \citet[][green five-pointed star] {Aguado2019}, with the theoretical Li-isochrones for the corresponding [Fe/H]=-5-8~dex models}. 
\label{li_teff_norris}
\end{figure}

We first focus on the [Fe/H] and \teff \ ranges of the Li plateau. We compare in Fig.~\ref{li_teff_norris} the observed Li abundances of normal Plateau dwarf stars as a function of their effective temperature to the corresponding theoretical Li isochrones in different [Fe/H] bins. Since some stars of the N23 sample have ages between 9 and 11~Gyr (see \S\ref{subsec:spins} for details on age determination), we show isochrones at two different ages, 12$\pm$1~Gyr and 10$\pm$1~Gyr for this sample. For the GALAH sample which contains mostly stars with ages above $\sim$ 11~Gyr, we only show the former one. Each metallicity bin extends by 0.25~dex to each side of the metallicity values of the grid. However, we made the lowest metallicity bins wider (both for the N23 and GALAH Pop~II samples) to include the star J0023$+$0307 with \feh=-5.8~dex (N23) and two stars that have \feh \ slightly below -2.75 (GALAH). Also, as well as in Fig.~\ref{hrd_norris}, the lowest metallicity bin shows models \feh=-5.8 with C- and Na-enhancement.

We see very good agreement between the models and the observations within 1$\sigma$ for most stars of both N23 and GALAH samples including the ones in the low-metallicity bins with -5.8$<$\feh$<$-2.75~dex. There are,  however, a couple of low metallicity dwarfs that sit below the theoretical plateau. Two of them have only lithium upper limit values. The third outlier is the most Fe-poor star of the N23 sample, J0023$+$0307. It shows no iron line but an upper limit for its \feh \ was derived from CaII~K  ($<$-5.8 and  $<$-6.6 from \citealt{Frebel2019} and \citealt{Aguado2019} respectively). It is marginally enriched in C for such an extremely Fe-poor star ([C/Fe]$>$3.9~dex), potentially excluding contamination from a binary companion. As discussed in the original papers, its abundance patterns including its overabundances in Mg and Si point towards a second-generation star formed out of material enriched by a single Population III massive star exploding as a fallback supernova.  While \citet{Frebel2019} and N23 derived values of 1.7 and 1.86 respectively for its A(Li), \citet{Aguado2019} derived a value of 2.02$\pm$0.08, much closer to the Plateau and our model predictions.

In Fig.~\ref{li_feh_dispersion} we show, for different ages,  the theoretical upper and lower Li envelopes as a function of [Fe/H]. These lines connect respectively the highest and lowest Li abundances at a given [Fe/H] and age predicted by the models in the \teff \ range of the plateau. As expected, the theoretical dispersion at a given age between the upper and lower Li envelopes is very small in the [Fe/H] range of the Spite plateau. Additionally, the envelopes become slightly positively sloped when age increases. Following N23, we computed the theoretical slope within the \feh \ interval of -3 to -1.5~dex, resulting in values of 0.04 and 0.05 at 10 and 13~Gyr respectively. At the same time, the observational slope among the normal sample stars (in the same metallicity range) is $0.08\pm0.02$ and $0.05\pm0.02$ for the N23 and combined N23+GALAH samples. This is in good agreement with \citealt{Asplund2006} who reports the value of $0.10\pm0.01$. The fact that the theoretical slope is even slightly lower than the observational one underscores the constancy of the $\log T_0$ value in the metallicity range of the Plateau. Otherwise, increasing turbulence would lead to a decrease in slope or even potential inversion. This imposes a strong constraint on the underlying mechanism, which is imitated in our models by parametric turbulence. Finally, Li depletion as modeled here does not predict a Li meltdown regime at very low metallicity, which further supports the non-stellar origin of this feature that seems to appear observationally only when CEMP stars are considered  (see \S~\ref{selection-field}). Contrary to the conclusion of \citet{Aguado2019}, we thus claim that the Li abundance in the most Fe-poor and not C-enhanced star, J0023+0307, which extends the Li plateau over more than five orders of magnitude in [Fe/H], strongly supports the stellar depletion solution to the cosmological Li problem. The determination of Li abundances in very metal-poor stars without (or with limited) C-enhancement is thus strongly encouraged.

\subsubsection{Comparison between models and observational data -- Li dispersion in more Fe-rich stars}
\label{subsec:comp_obs_data_highZ}

As discussed above (see also \S\ref{subsec:robustness}), the stellar Li-depletion solution to the Li plateau constancy is sustained by the fact that very Fe-poor low-mass stars have very similar compact structures on the main sequence. Stellar compactness decreases however around [Fe/H]$\sim$-1.5~dex, inducing larger variations in the temperature and density gradients in the stellar radiative zones as the stellar mass and metallicity vary. In addition, while at low metallicities we see only old stars, stellar populations in the more Fe-rich domain of the Milky Way have high dispersion in ages. These two facts are expected to result in a high dispersion of Li abundance in more Fe-rich dwarfs because the Li depletion rate depends on \feh, \ mass, and age. Stellar lifetime is also highly dependent on mass and, to a much lesser extent, on metallicity which leads to the fact that the upper mass limit decreases when we look at older stars. 

\begin{figure}
\centering
\includegraphics[width=1\linewidth]{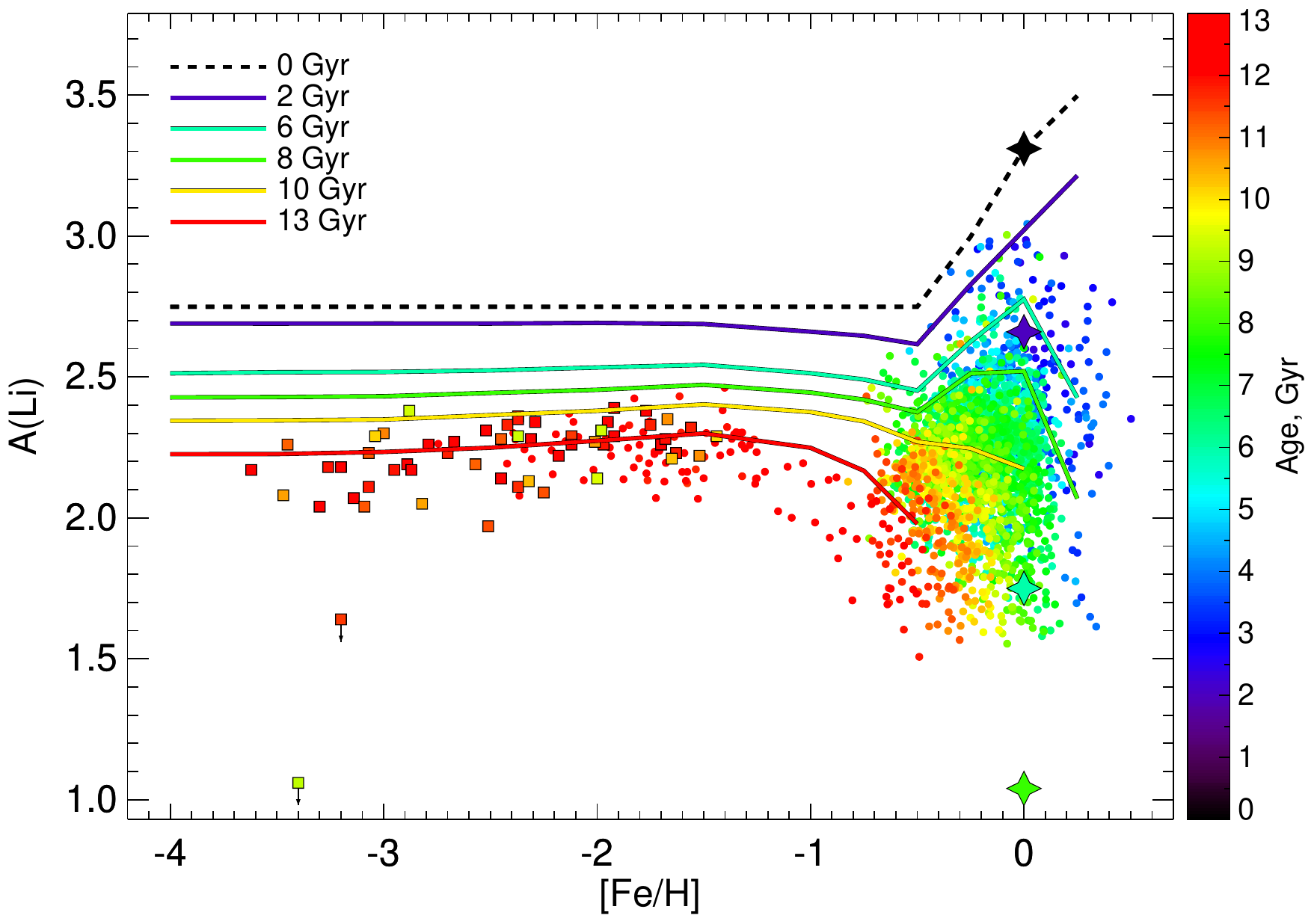}
\includegraphics[width=1\linewidth]{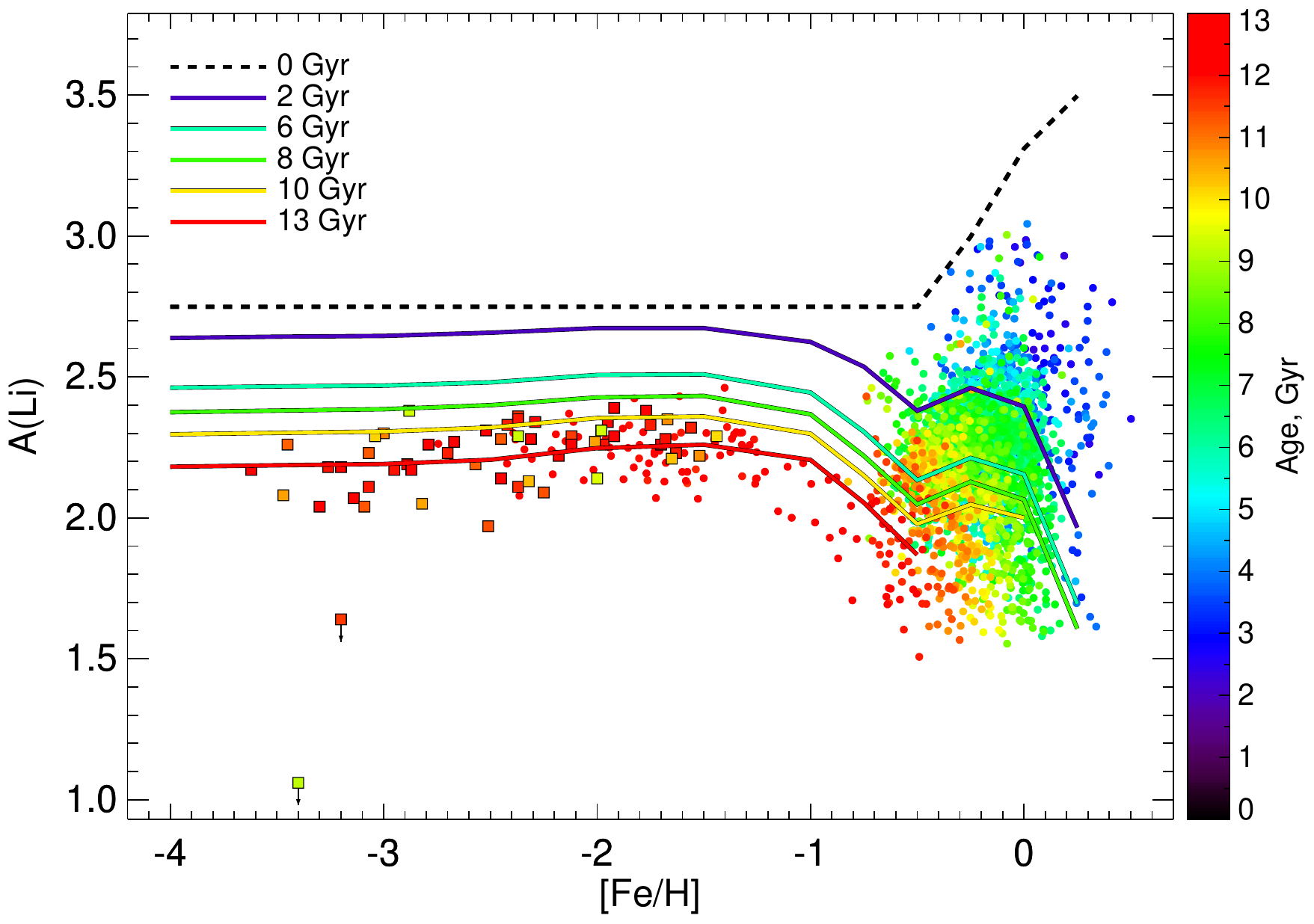}
\caption{Li abundance (3D,NLTE) as a function of \feh \ for the N23 and GALAH~DR3 sample stars (squares and circles respectively) with age color-coded. Solid lines show the theoretical Li upper and lower envelopes (upper and bottom panels respectively)  at different ages (see the legend), i.e., the maximum and minimum Li values at a given [Fe/H] and age expected from the models with 5800~K$\lesssim$\teffzams$\lesssim$6500~K. 
We do not show it here, but the highest/ lowest A(Li) values for the models at [Fe/H]=-5.8~dex are 2.34/2.29 and 2.23/2.18 at the ages of 10 and 13~Gyr respectively. The dashed line shows the initial Li abundance assumed in the models. In the upper panel, the four-pointed stars show predictions from the models with log$T_0$=6.42 and solar metallicity from D21b with age similarly colour-coded.}
\label{li_feh_dispersion}
\end{figure}

The corresponding theoretical behaviour of the upper and lower Li envelopes (hence of the expected dispersion) at different ages is shown and compared to the observations in Fig.~\ref{li_feh_dispersion}. The data correspond to the normal dwarf stars in the \teff \ range from 5800~K to 6500~K 
from N23 and the extended sample of GALAH~DR3 with ages colour-coded (the points are plotted in the age ascending order to show the oldest stars that exist in different parts of this diagram). The dashed line shows the initial Li value assumed in our models as a function of [Fe/H] (see \S\ref{subsec:input_physics}). Although the linear interpolation we used between the BBN and meteoritic Li values is not fully justified and might cause discrepancies between the observed Li abundances and model predictions, it follows the observational trend in the youngest and most Fe-rich stars, and it cannot affect the theoretical Li dispersion pattern qualitatively. At the metallicities higher than that of the Li plateau, the models predict increasing Li dispersion and fit very well the observed upper and lower Li envelopes at different ages. However, there are still many stars with [Fe/H] above $\sim-0.5$~dex and ages higher than $\sim$~6~Gyrs that present Li values below the theoretical predictions. This could be due to the fact that our models are computed only for $\log T_0$=6.28. Indeed and as discussed in \S\ref{subsec:turbulence_calib}, stars with [Fe/H] above $\sim -1.5$~dex require an increase of this parameter, which would lead to an increase in Li depletion rate and, hence, a decrease of Li abundance. For this purpose, we also show upper values at the solar metallicity predicted by models for dwarfs with $\log T_0$=6.42 and \teffzams$<$6500~K from D21b for which turbulence was adjusted to reproduce solar-type stars around solar metallicity. Since these models are computed with a mass step of 0.1\ms, for higher precision, we interpolated the Li values from the models at ages 2, 6, and 8~Gyr. Interpolated values approximately correspond to masses of 1.25, 1.15, and 1.05\ms. These results shall help future studies of the physical origin of the process that we simulate as parametric turbulence. 

\subsection{Globular cluster stars - Lithium and metals}
\label{subsec:selection-GC}

The abundance variations of metals around the MS turnoff in globular clusters have been successfully used to pinpoint the combined effects of atomic diffusion and turbulence in Pop~II low-mass stars \citep[e.g.][]{Korn2007, Lind2008, Nordlander2012, Gruyters2013, Gruyters2014,Gavel2021,Nordlander2024}. We have carried out this additional test on our models using Mg and Li data for the globular cluster NGC~6752 from \citet{Gruyters2013} and \citet{Gruyters2014} (hereafter G13 and G14 correspondingly), where the former one provides measurements for individual post-turnoff stars and the latter one averaged values for four groups of stars at different evolutionary stages: turnoff, SGB, base of the RGB, and RGB. Although the G14 sample is more complete, the G13 sample covers a wider \teff \ range. The G14 sample contains 146 and 124 stars with measurements of Mg and Li abundance correspondingly. To avoid initial abundance variations among the multiple stellar populations hosted by the cluster \citep[for a review on multiple stellar populations in globular clusters, see ][]{2019A&ARv..27....8G}, we consider only those stars from G14 that were identified as being part of the ``primordial'' (i.e., first) population. These stars should have the same original composition as field halo stars of similar \feh. We are left with 25 and 34 stars with measurements of Li and Mg respectively. We adopt the same age for the cluster as G14 (13.5~Gyr). The authors derive [Fe/H]=-1.6 and \afe=0.3 for the cluster. Other estimations suggest its \feh \ value is around -1.5 (e.g. \feh=-1.49, \citealt{Souza2020}; -1.48, \citealt{Gratton2005}; -1.55, \citealt{Harris1996}). Considering these slight differences and in order to save computational time, we used the already computed models for \feh=-1.5 and \afe=0.3. We completed them with a few additional models around the cluster turnoff mass of 0.8\ms\footnote{To this end, we computed models with initial masses of 0.795, 0.798, 0.802, 0.805, 0.808, and 0.81\ms.} to trace the SGB and RGB with a higher level of discreteness. 

\begin{figure}
\centering
\includegraphics[width=1\linewidth]{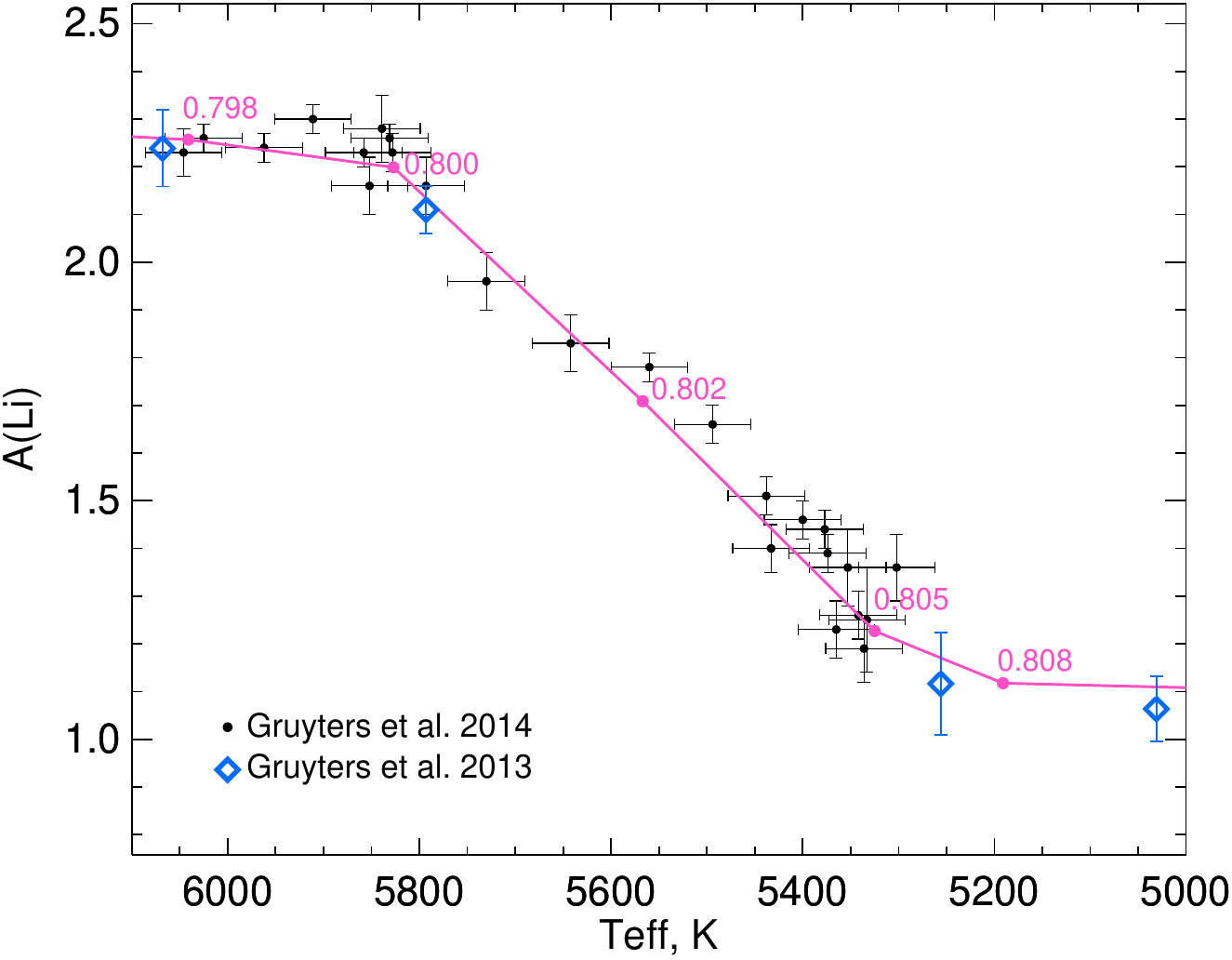}
\includegraphics[width=1\linewidth]{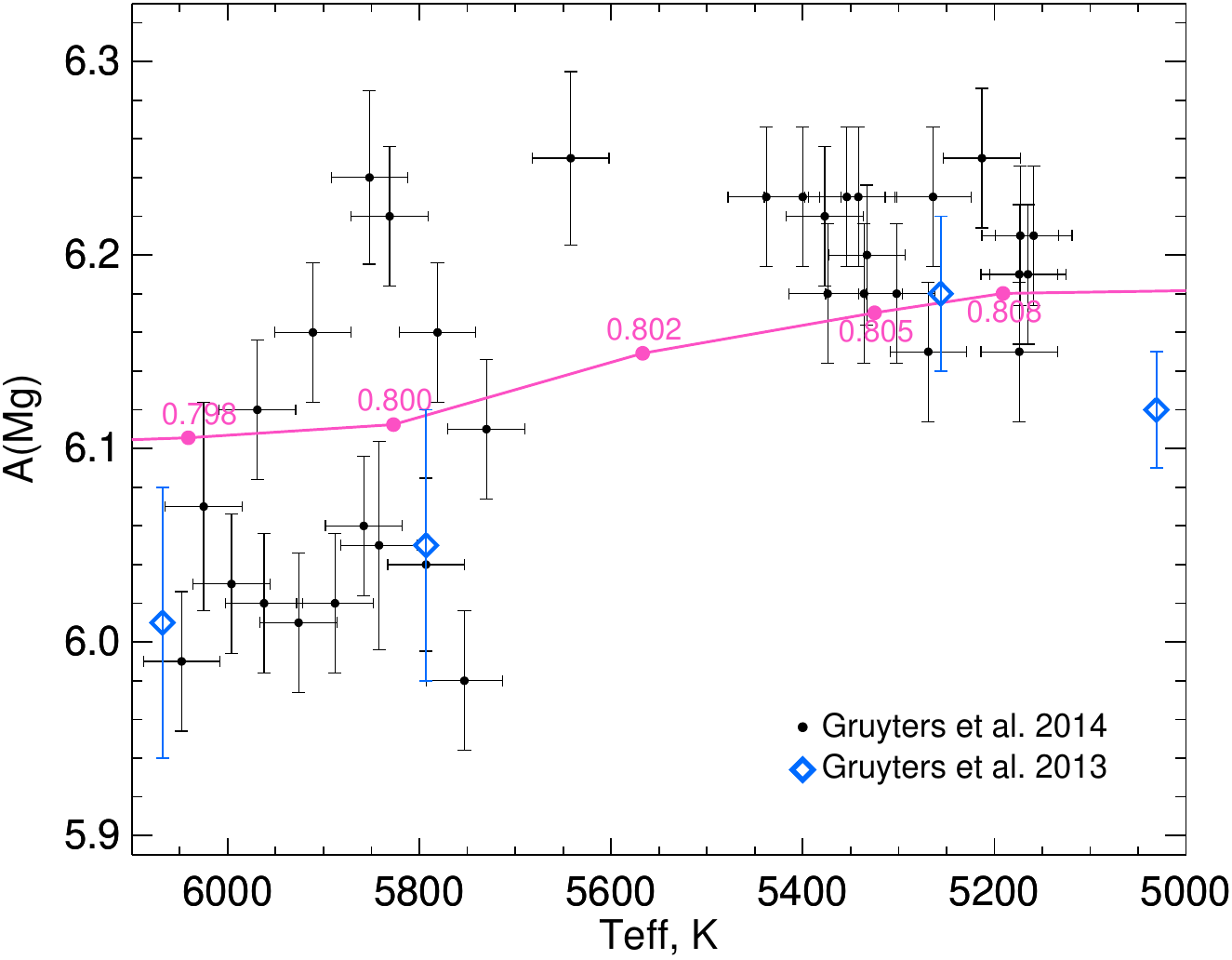}
\caption{\textit{Upper panel:} trend of Li abundance in stars of the primordial population of NGC~6752. Black points correspond to the data points from G14 and blue diamonds show the averaged values for different evolutionary stages of stars from G13 (turnoff, SGB, base of the RGB, RGB from left to right). The model prediction at 13.5~Gyr is shown with the magenta line with points labeled according to their mass. \textit{Bottom panel:} The same, but for the abundance of Mg.} 
\label{gc_abund}
\end{figure}

We compare the model predictions to the Li and Mg observations as a function of \teff \ (as a proxy for the evolution stage) in Fig.~\ref{gc_abund}. For both elements, the magenta lines connect the theoretical abundances in models of different initial masses at 13.5~Gyr (i.e., isochrones for Li and Mg).  
Although we calibrated models to reproduce Li abundance in turnoff MS stars, we obtain a very good agreement between the observational data and model for Li beyond the MS until the end of the first dredge-up. As discussed in \S\ref{subsec:spins}, the abundances of heavier elements (including Mg) that have accumulated below the convective envelope along the MS due to atomic diffusion recover their initial values as the first dredge-up occurs (see also Fig.~\ref{moh_li_kippen} for the global metallicity [M/H]). The theoretical predictions for Mg were homogeneously shifted by -0.14~dex to account for the differences between the initial Mg content assumed in our models and the observations. The shift value was computed to minimize the $\chi^2$ discrepancy between the model and observation. Although the models were not calibrated to fit the Mg abundance increase in post-turnoff stars, they still reproduce the observed trend at a fairly high level of confidence. Our results are in agreement with those of the models of G14 with atomic diffusion and parametric turbulence derived a slightly lower value of log~T$_0$=6.20 than ours for parametric turbulence in order to recover the main observational trends for metals in NGC~6752. The corresponding initial Li abundances they retrieved with their optimal model are thus slightly lower (A(Li)=2.58$\pm$0.10) than the BBN+CMB value we assume in our models with log~T$_0$=6.28. We consider, however, that the difference is not compiling, for two reasons. First, the different input microphysics in G14 and our models can lead to slight stellar structure differences, with differences in the optimal log~T$_0$ simply hiding our ignorance about the origin of the parametric turbulence. Second, the significance of the trends for metals is relatively weak in NGC~6752. 

\section{Summary and conclusions}
\label{sec:summary}
The discrepancy between the Li value predicted by BBN+CMB and the one observed along the Spite plateau has long been a matter of debate. In this work, we investigated the stellar depletion solution using the same assumptions as in the models of \citet{Dumont2021a,Dumont2021b} that explain the  Li behaviour and the surface and internal rotation of Population~I low-mass stars. They include rotation-induced processes, parametric vertical viscosity, atomic diffusion, penetrative convection, and parametric turbulence. Given the lack of asteroseismic data for Pop~II Plateau dwarf stars, we adopted the same value for the vertical viscosity as required in the $_{\nu}R1^{T_0}_A$ models of D21a/b to explain the internal rotation profile of solar-type stars, and we show that variations around this value do not significantly impact the Li content at the age of Pop~II stars (see \S\ref{subsec:viscosity}). We adopted the same initial rotation periods as  \citet{Amard2019} and D21a,b for Pop~I low-mass stars and show that Li depletion at the age of the Plateau dwarfs is marginally impacted by the initial rotation rate. Finally, the only difference to the $_{\nu}R1^{T_0}_A$ models of D21a/b is the value of $\log T_0$ where the efficiency of parametric turbulence is anchored, which we adjusted to counteract atomic diffusion and fit the surface abundance of Li observed along the Plateau. Using the BBN+CMB Li value as the initial Li abundance for Pop~II stars, we found that the value of $\log T_0$=6.28 fits the best the today-observed value of A(Li)=2.3. This value is in good agreement with $\log T_0$=6.25 found by \citet{Richard2005} whose models include only atomic diffusion and parametric turbulence that turn out to be the dominant processes for surface Li depletion along the plateau compared to rotation-induced meridional circulation and horizontal shear as modeled in the $_{\nu}R1^{T_0}_A$  configuration.

We have demonstrated that the constancy of the Li Spite plateau over a large range of metallicities results naturally from the similar compactness of the plateau dwarf stars, whose structure does not significantly change in a wide metallicity range. Consequently, these stars are expected to undergo the combined action of parametric turbulence and atomic diffusion at a similar level of efficiency.

At the same time, we have explained the constancy of the Spite plateau in quite a wide range of \teff. Although, at a given metallicity, atomic diffusion efficiency increases with \teff, its effect is balanced by the changing of the density gradient below the base of the convective envelope. Eventually, these two effects create a domain in the [Fe/H]-\teff \ space where stars have very similar rates of Li depletion, which results in a plateau with very little Li dispersion. When using the same $T_0$ value over a [Fe/H] range covering five orders of magnitude below -1.5~dex, the $_{\nu}R1^{T_0}_A$ models predict a very modest positive slope for the Li plateau, but no  Li meltdown at [Fe/H] below -2.5~dex. This is in excellent agreement with the trend found after a detailed analysis of the peculiarities of dwarf stars with Li data in the very low metallicity regime. There, the dwarf stars that lie below the plateau are all CEMP or other chemically peculiar stars, supporting the environmental origin (influence of Pop~III stars in the original composition, and/or binarity-induced processes) of the so-called meltdown. The Li abundance close to the Plateau of the most Fe-poor star, J0023+0307, which is not C-enriched, provides strong additional support to the stellar depletion solution of the cosmological problem.

We compared trends of Li and Mg in post-turnoff stars of the globular cluster NGC~6752 with the model predictions. These stars experience the first dredge-up caused by the rapid deepening of the convective envelope which results in the increase of most metals and the decrease of Li. We found a perfect agreement between observed Li abundances and models that have the most recent measurement of the primordial Li abundance of A(Li)=2.75~dex \citep{Pitrou2018,Coc2017} as a starting point. Also, the modeled trend for Mg abundance fits well with the observational data. This further supports the stellar depletion solution to the cosmological Li problem. Possibly,  Mg and Li data in post-turnoff stars in metal-rich globular clusters could help estimate how the Li abundance increased in the interstellar matter from the BBN to the meteoretic value in the -1$<$\feh$<$-0.5~dex range. Up to date, it is in this range that there are the greatest uncertainties in the behaviour of Li. Understanding this may help shed light on the history of enrichment in the Galaxy.

The same stellar structure considerations consistently explain the change of Li depletion and dispersion regime for [Fe/H] around -1.5~dex, i.e., at the transition in metallicity between Pop~II to Pop~I stars. The stellar compactness decreases strongly above this metallicity, and the density at the base of the convective envelope (hence the atomic diffusion efficiency) varies strongly with \teff \ and metallicity. This explains the need for a higher anchoring temperature for parametric turbulence for Pop~I dwarfs than for Pop~II stars along the plateau. This also leads naturally to higher Li dispersion in Pop~I stars at a given age (e.g., $\sim$0.6-0.7~dex at the solar metallicity at 2~Gyr), and more generally since Pop~I low-mass stars that cover a large range in age. This plateau-to-scatter transition pattern is in good agreement with the observational data, which further supports a consistent stellar depletion solution to the Li behaviour in low-mass stars from different Galactic populations and to the Li cosmological problem.  

\begin{acknowledgements}
SB and CC acknowledge financial support from the Swiss National Science Foundation (SNF; Project 200021$-$212160). TD is supported by the European Union, ChETEC-INFRA, project no. 101008324. We thank John Norris for kindly providing data on stellar parameters and abundances. This research has made use of NASA’s Astrophysics Data System Bibliographic Services. We extensively used the \textsc{SIMBAD} database \citep{simbad}, VizieR catalog access tool \citep{vizier}, and \textsc{TOPCAT} \citep{topcat}.
\end{acknowledgements}

\bibliographystyle{aa}
\bibliography{pop_II}

\end{document}